%% file: Main.tex
\documentclass[12pt,a4]{article}
\usepackage{amsfonts}
\usepackage{amsmath}
\usepackage{amssymb}
\usepackage{amsbsy}
\usepackage{color}
\usepackage{graphicx}
\usepackage{url}
\usepackage{sectsty}
\usepackage{appendix}
\usepackage{aicescover}
\usepackage{cite}
\usepackage{tensor} 
\usepackage{pgf}
\usepackage{tikz}
\usetikzlibrary{shapes,snakes,arrows}

\usepackage{subfigure}

\input{symacros}
\begin{document}
%

\title{Towards an Efficient Use of the BLAS Library for Multilinear Tensor Contractions}






\author{%
  Edoardo Di Napoli
  \footnote{J\"ulich Supercomputing Centre, Institute for Advanced
    Simulation, Forschungszentrum J\"ulich, Wilhelm-Johnen strasse,
    52425--J\"ulich, Germany.  {\tt e.di.napoli@fz-juelich.de}.} \and
  Diego Fabregat-Traver%
  \footnote{AICES, RWTH-Aachen University, 52056--Aachen, Germany.
    {\tt \{fabregat,pauldj\}@aices.rwth-aachen.de}.} \and Gregorio
  Quintana-Ort\'i%
  \footnote{Depto. de Ingenier\'{\i}a y Ciencia de Computadores,
    Universidad Jaume I, 12.071--Castell\'on, Spain.  {\tt
      gquintan@icc.uji.es}.} \and Paolo Bientinesi$^\dagger$ }
\aicescoverack{Financial support from the \\ Deutsche
  Forschungsgemeinschaft (German Research Foundation) through grant
  GSC 111 and the VolkswagenStiftung through the fellowship
  “Computational Sciences” is gratefully acknowledged.}
\aicescoverpage
\maketitle

\begin{abstract} 
  Mathematical operators whose transformation rules constitute the
  building blocks of a multi-linear algebra are widely used in physics
  and engineering applications where they are very often represented
  as tensors. In the last century, thanks to the advances in tensor
  calculus, it was possible to uncover new research fields and make
  remarkable progress in the existing ones, from electromagnetism to
  the dynamics of fluids and from the mechanics of rigid bodies to
  quantum mechanics of many atoms.  By now, the formal mathematical
  and geometrical properties of tensors are well defined and
  understood; conversely, in the context of scientific and
  high-performance computing, many tensor-related problems are still
  open.  In this paper, we address the problem of efficiently
  computing contractions among two tensors of arbitrary dimension by
  using kernels from the highly optimized BLAS library.  In
  particular, we establish precise conditions to determine if and when
  \gem, the kernel for matrix products, can be used.  Such conditions
  take into consideration both the nature of the operation and the
  storage scheme of the tensors, and induce a classification of the
  contractions into three groups. For each group, we provide a recipe
  to guide the users towards the most effective use of BLAS.
\end{abstract}



\input{body}

\section*{Acknowledgements}
Financial support  
from the Deutsche Forschungsgemeinschaft (German Research Association)
through grant GSC 111 is gratefully acknowledged.
This research was in part supported by the VolkswagenStiftung through
the fellowship ``Computational Sciences''.

\bibliographystyle{plain}

\end{document}

%% file: symacros.tex
\newcommand\be{\begin{equation}}
\newcommand\ee{\end{equation}}
\newcommand\bea{\begin{eqnarray}}
\newcommand\eea{\end{eqnarray}}
\newcommand\bi{\begin{itemize}}
\newcommand\ei{\end{itemize}}
\newcommand\nn{\nonumber}

\newcommand\sv{$\vec{s}$}

\newcommand{\gem}{{\sc\small GEMM}}
\newcommand{\copgem}{{\sc\small COPY + GEMM}}
\newcommand{\copgemv}{{\sc\small COPY + GEMV}}
\newcommand{\gemv}{{\sc\small GEMV}}
\newcommand{\ger}{{\sc\small GER}}
\newcommand{\dt}{{\sc\small DOT}}
\newcommand{\tes}{{\sc tesla}}
\newcommand{\pec}{{\sc peco}}

\newcommand\eqalign[1]{%
	\vcenter{%
		\normalbaselines \advance\baselineskip 5pt
		\advance\lineskip 5pt \tabskip=0pt
		\halign{%
			&\hfil $\displaystyle{##{}}$&
			$\displaystyle{{}##}$\hfil\cr
			#1\crcr
			}%
		}%
	}

\newcommand\dotline{\par\hbox to \hsize{\dotfill}\par}

\makeatletter
\def\befored@t#1.#2.#3;{#1}
\def\afterd@t#1.#2.#3;{#2}
\def\refhead#1{\edef\next{\ref{#1}}\expandafter\befored@t\next..;}
\def\reftail#1{\edef\next{\ref{#1}}\expandafter\afterd@t\next..;}
\def\lsim{\mathrel{\mathpalette\@versim<}}
\def\gsim{\mathrel{\mathpalette\@versim>}}
\def\@versim#1#2{\vcenter{\offinterlineskip
        \ialign{$\m@th#1\hfil##\hfil$\crcr#2\crcr\sim\crcr } }}
\newcommand\becomes[1]{\mathchoice{\becomes@\scriptstyle{#1}}
   {\becomes@\scriptstyle{#1}} {\becomes@\scriptscriptstyle{#1}}
   {\becomes@\scriptscriptstyle{#1}}}
\def\becomes@#1#2{\mathrel{\setbox0=\hbox{$\m@th #1{\,#2\,}$}%
        \mathop{\hbox to \wd0 {\rightarrowfill}}\limits_{#2}}}
\makeatother

%% file: body.tex
\section{Introduction}
In the last decade, progress in many scientific disciplines like
Quantum Chemistry~\cite{CCD} and General
Relativity~\cite{Carroll} was based on the ability of simulating
increasingly large physical systems.  Very often, the simulation codes
rely on the contraction of matrices and
tensors\footnote{ A contraction is the
  generalization of the matrix product; see Sec.~\ref{sec:sec12}. }~\cite{COD};
surprisingly, while matrix operations are supported by numerous highly
optimized libraries, the domain of tensors is still sparsely
populated.  In fact, the contrast is staggering: On the one hand,
\gem---the matrix-matrix multiplication kernel from the BLAS
library\footnote{The Basic Linear Algebra Subroutines (BLAS) are a
  collection of common building blocks for linear algebra
  calculations.}---is possibly the most optimized and used routine in
scientific computing; on the other hand, high-performance black-box
libraries for high dimensional contractions are only recently starting
to appear.  This paper attempts to fill the gap, not by
developing an actual tensor library---a ``tensor-BLAS''---but by
identifying when and how contractions among tensors of arbitrary
dimension can be expressed, and therefore computed, in terms of the
existing highly-efficient BLAS kernels.

Since the introduction of BLAS 1 in the early 70s'~\cite{BLAS1},
the numerical linear algebra community has made tremendous progress in
the generation of high-performance libraries. Nowadays BLAS
consists of three levels, corresponding to routines for vector-vector,
matrix-vector and matrix-matrix operations~\cite{BLAS2,BLAS3}.
From a mathematical perspective, it might appear that this structure
introduces unnecessary duplication: For instance, 
a matrix-matrix multiplication (a level 3 routine)
can be expressed in terms of matrix-vector products (level 2), 
which in turn can be expressed in terms of inner products (level 1).
Beyond convenience, the layered structure is motivated by the efficiency of the routines:
Due to the increasing
ratio between operation count and number of memory accesses---about 
1/2, 2 and $n$/2 for BLAS 1, 2, and 3, respectively---,
higher-level operations offer better opportunity to amortize the costly
memory accesses with calculations.  In practice, this means that level 3
routines attain the best performance and should be preferred whenever
possible.

Intuitively, one would expect the ratio between flop count and memory
accesses to continue growing with the dimension of the tensors
involved in the contraction, suggesting a level 4 BLAS and beyond.
While it might be worthwhile to further explore this research idea,
since the potential of today's processors is already fully exploited
by BLAS 3 kernels, we aim at expressing arbitrary contractions in
terms of those kernels.

Our study deals with contractions of any order between two tensors of
arbitrary dimension; the objective is to provide computational scientists with
clear directives to make the most out of the BLAS library, thus attaining
near-to-perfect efficiency. 

This paper makes two main contributions. First, it contains a set of
requirements to quickly determine if and how a given contraction can
be computed utilizing \gem, the best performing BLAS kernel.  These
requirements lead to a 3-way classification of the contractions
between two generic tensors. The second contribution consists of
guidelines, given in the form of recipes, to achieve an optimal use of
BLAS for each of the three classes, even when \gem\ is not directly
usable.  In some cases, the use of BLAS 3 is only feasible in
combination with copying operations, effectively equivalent to a
transposition; in such cases, our guidelines indicate how the tensors
should be generated so that the transposition is altogether avoided.
To support our claims, we performed a number of experiments with
contractions that arise frequently in actual applications; in all
cases, our guidelines are confirmed by the observations.  Furthermore,
since our analysis of contractions from actual applications provides
examples of operations that arise frequently and that are not
currently supported by BLAS, we believe that this study will be
helpful to library developers who wish to support tensor operations.

\paragraph{Related work} 
In recent years, the problem of efficiently handling tensor operations
has been increasingly studied.  A multitude of methods were proposed
to construct low dimensional approximations to tensors.  This research
direction is very active, and orthogonal to the techniques discussed
in this paper; we refer the reader to~\cite{KB} for a review, and
to~\cite{TT,PAR} for other advances.  Two projects focus specifically
on the computation of contractions: the Tensor Contraction Engine
(TCE) and Cyclops~\cite{TCE1,CTF2}.  TCE offers a domain-specific
language to specify multi-index contractions among two or more
tensors; through compiler techniques it then returns optimized Fortran
code under given memory constraints~\cite{TCE2,TCE3}.  Cyclops is
instead a library for distributed memory architectures that targets
high dimensional tensors arising in quantum chemistry~\cite{CTF1};
high-performance is attained thanks to a communication-optimal
algorithm, and cyclic decomposition for symmetric tensors.  We stress
that the approach described in this paper goes in a different
direction, targeting dense tensors, and decomposing the contractions
in terms of standard matrix kernels.

\paragraph{Organization of the paper}  
The paper is organized as follows. In Sec.~\ref{sec:sec2}, we
introduce definitions, illustrate our methodology through a basic
example, and present the tools and notation used in the rest of the
paper.  Section \ref{sec:sec3} constitutes the core of the paper: It
contains the requirements for the use of \gem, the classification of
contractions, and recipes for each of the contraction classes.  Next,
in Sec.~\ref{sec:sec5} we present a series of examples with
performance results. Concluding remarks and future work are discussed
in Sec.~\ref{sec:conclusions}.


\section{Preliminaries}
\label{sec:sec2}

We start with a list of definitions for acronyms and concepts 
used throughout the paper.
We then illustrate our approach by means of a simple example in which 
the concept of ``slicing'' is introduced. 
Once a formal definition of algebraic tensor operations is given, we conclude
by showing how slicing is applicable to contractions between 
arbitrary tensors.

\subsection{BLAS definitions and properties}
\label{sec:blasdef}
The main motivations behind the Basic Linear Algebra Subroutines were
modularity, efficiency, and portability through
standardization~\cite{DodsonLewis}.
With the advent of architectures with a
hierarchical memory, BLAS was extended with its levels 2 and 3,
raising the granularity level to allow an efficient use of the 
memory hierarchy---by amortizing the costly data movement with 
the calculation of a larger number of floating point operations.
With respect to the full potential of a processor, the
efficiency of BLAS 1, 2, and 3 routines is roughly 5\%, 20\% and 90+\%,
respectively.  It is widely accepted that for large enough operands, these
values are nearly perfect.
Moreover, the scalability of BLAS 1 and 2 kernels is rather limited,
while that of BLAS 3 is typically close to perfect.
As demonstrated by its universal usage, BLAS succeeded in
providing a portability layer between the computing architecture and both
numerical libraries and simulation codes.

In addition to the reference library,\footnote{Available at {\tt http://www.netlib.org/blas/.}}
nowadays many implementations of BLAS exist, including hand-tuned~\cite{1377607,Goto1},
automatically-tuned~\cite{atlas-sc98,Whaley:PhD}, and versions developed by processors
manufacturers~\cite{MKL1,ESSL,ACML}.  Despite their excellent efficiency, such
libraries might not be well suited for tensor computations. 
For instance, in the
following it will become apparent that the design of the BLAS interface falls
short when matrices are originated by slicing higher dimensional objects;
specifically, costly transpositions are caused by the constraint that matrices
have to be stored so that one of the dimensions is contiguous in
memory. Recent efforts aim at lifting some of these
shortcomings~\cite{BLIS}.

A list of BLAS related definitions follows.

\begin{itemize}

\item {\sc Stride}: 
  distance in memory between two adjacent elements of a vector or a matrix.
  BLAS requires matrices to be stored by columns, that is, with vertical $stride =
  1$.

\item {\sc Leading dimension}:
  distance between two adjacent row-elements in a matrix stored by columns. 
  In BLAS routines, leading dimensions are specified through 
  arguments. 

\item {\gem}: $ C \leftarrow \alpha  A^\bullet B^\bullet + \beta C$;
  kernel for general matrix-matrix multiplication (BLAS 3). 
  $A, B$ and $C$ are matrices; $\alpha$ and $\beta$ are scalars;
  $M^\bullet$ indicates that matrix $M$ can be used with/without transposition. 

\item {\gemv}: $ y \leftarrow \alpha A^\bullet x + \beta y$;
  kernel for general matrix-vector multiplication (BLAS 2).
  $A$ is a matrix, $x$ and $y$ are vectors, and $\alpha$ and $\beta$ are 
  scalars;  $A^\bullet$ indicates that matrix $A$ can be used with/without transposition. 

\item {\ger}: $ C \leftarrow \alpha x y^H + C$;
  outer product (BLAS 2).
  $C$ is a matrix, $x$ and $y$ are vectors, and $\alpha$ is a scalar.

\item {\dt}: $ \alpha \leftarrow x^H y$;
  inner product (BLAS 1).
  $x$ and $y$ are vectors, and $\alpha$ is a scalar.

\end{itemize}

\subsection{A first example}
\label{sec:example}

In order to illustrate our methodology with a first well-known example,
and thus facilitate the following discussion, here we consider one of the most basic
contractions, the multiplication of matrices.

To the reader familiar with the jargon based on Einstein convention, 
given two 2-dimensional tensors $A^{ih}$ and $B_{hj}$, we want to contract
the $h$ index.  In linear algebra terms, the
matrices $A$ and $B$ have to be multiplied together.  
\[
C^i_j := A^{ih} B_{hj}, 
\text{ \quad or equivalently: \quad } 
c_{ij} := \sum_h a_{ih} b_{hj}, \ \forall i \forall j  
\]
For the sake of this example, we assume that only level 1 and 2 BLAS
are available, while level 3 is not; in other words, we assume that
there exists no routine that takes (pointers to) $A$ and $B$ as input,
and directly returns the matrix $C$ as output. 
Therefore, we decompose the operation in terms of level 1
and 2 kernels by slicing the operands,
exactly the same way we will later decompose a
contraction between tensors in terms of matrix-matrix multiplications.

In the following, we provide a number
of alternatives for slicing the matrix product. Matrices 
$A, B$ and $C$ are of size $m\times k$,
$k\times n$, and $m\times n$, respectively. Also, $F^i$ and $F_j$ 
respectively denote the $i$-th row and $j$-th column of the matrix $F$. 

\begin{enumerate}
\item {\bf $A$ is sliced horizontally.} 
  The slicing of $i$, the first index of $A$, leads to a decomposition of 
  $C$ as a sequence of independent \gemv s:
  \fbox{$C := \forall_{i} \ a^i B$}
  \[
  C := A B = 
  \left[ 
  \begin{array}{c}
    a^1 \\ \hline
    \vdots \\ \hline
    a^m
  \end{array}
  \right]
  B
  =
  \left[ 
  \begin{array}{c}
    a^1 B \\ \hline
    \vdots \\ \hline
    a^m B
  \end{array}
  \right].
  \]

\item {\bf $B$ is sliced vertically.} 
  The slicing of $j$, the second index of $B$, also leads to a decomposition of 
  $C$ as a sequence of independent \gemv s:   
  \fbox{$C := \forall_j \ A b_j$}
  \[
  C := A B = A [ b_1 | b_2 | \dots | b_n ] =
  [ A b_1 | A b_2 | \dots | A b_n ].
  \]

\item {\bf $A$ is sliced vertically and $B$ horizontally.}
  The slicing of $h$, the contracted index,  
  leads to expressing $C$ as a sum of \ger s:
  \fbox{$C := \sum_h a_h b^h$}
  \[
  C := A B = 
  [ a_1 |  \dots | a_k ] 
  \left[ 
  \begin{array}{c}
    b^1 \\ \hline
    \vdots \\ \hline
    b^k
  \end{array}
  \right]
  =
  a_1 b^1 + a_2 b^2 + \dots + a_k b^k.
  \]

\item {\bf $A$ is sliced horizontally and $B$ vertically.}
  If both indices $i$ and $j$ are sliced, 
  $C$ is decomposed as a two-dimensional
  sequence of \dt s:
  \fbox{$ C := \forall_i \forall_j \ a^i b_j$}
  \[
  C := A B = 
  \left[ 
  \begin{array}{c}
    a^1 \\ \hline
    \vdots \\ \hline
    a^m
  \end{array}
  \right]
  [ b_1 | \dots | b_n ] =
  \left[ 
  \begin{array}{c | c | c}
    a^1 b_1 & \dots & a^1 b_n \\ \hline
    \vdots  & \ddots &  \\ \hline
    a^m b_1 &        & a^m b_n
  \end{array}
  \right].
  \]

\end{enumerate}

While mathematically these four expressions compute exactly the same
matrix product, in terms of performance they might differ substantially. From
this perspective, slicings 1--3 lead to level 2 BLAS kernels (GEMVs for 1 and
2, GERs for 3) and are thus to be preferred to slicing 4, which instead leads
to level 1 kernels (DOT).  In the following sections we will follow the same
procedure for higher-dimensional tensors and multiple contractions.

\subsection{Tensors definitions and properties}
\label{sec:sec12}

Due to their high-dimensionality, tensors are usually presented together with
their indices.  As commonly done in physics related literature, in the rest of
the paper we use greek letters to denote indices. We now illustrate some
concepts which are not strictly necessary to understand the methodology
introduced in the following sections, but that are important to establish a
connection with the language commonly used by physicists and chemists.

As explained in Appendix~\ref{Appendix}, there is a substantial
difference between upper (contravariant) and lower (covariant)
indices. In general, the connection between covariant and
contravariant indices is given by a special tensor $g_{\mu \nu}$
called ``metric''.  For example, an upper index of a vector $x$ can be
lowered by multiplying the vector with the metric tensor (in itself
having properties similar to those of a symmetric matrix): 
\be
x_{\mu'} = \sum_{\mu = 1}^N x^\mu g_{\mu \mu'} \equiv x^\mu g_{\mu
  \mu'}. \nn 
\ee 
Notice that the right hand side is written without
explicit reference to the sum over the $\mu$ index. This convention is
usually known as the ``Einstein notation'': Repeated indices come in
couples (one upper and one lower index) and are summed over their
entire range. We refer to this operation as a ``contraction''. It is
important to keep in mind that contractions happen only between equal
indices lying on opposite positions.  Following this simple rule,
matrix-vector multiplication is written as
\be 
Ax \equiv A^{\alpha \beta} g_{\beta
  \gamma} x^\gamma \equiv A\indices{^\alpha_\gamma} x^\gamma. \nn 
\ee
Since a tensor can have more than one upper or lower index, we
interpret lower indices as generalized ``column modes'' and the upper
indices as generalized ``row modes''.  In general, it is common to
refer to a $k$-dimensional tensor having $i$ row modes and $j$ column
modes (with $i + j = k$) as a ($i,j$)-tensor or an ``order-$k$
tensor''.  Finally, since relative order among indices is important,
the position of an index should be always accounted for; this is of
particular importance in light of the results we present.

Having established the basic notations and conventions (see
Appendix~\ref{Appendix} for more details), we are ready to study
numerical multilinear operations between tensors.
We begin by looking at an example where only one index ($\gamma$) is
contracted. In the language of linear algebra, the operation can be
described as an inner product between a (3,0)-tensor and a (2,0)-tensor 
\be
R\indices{^{\alpha \beta \rho}} := 
(T\indices{^{\alpha \beta \gamma}}) \cdot (S^{\gamma \rho}).
\label{eq:tensdot}
\ee
In accordance with the convention, 
the summation over $\gamma$ can only take place once 
the index is lowered in one of the two tensors. 
For example, $\gamma$ can be lowered in $T$ as
\be
T\indices{^{\alpha \beta \gamma}} \longrightarrow  T\indices{^{\alpha \beta \mu}} g_{\mu \gamma} \equiv T\indices{^{\alpha \beta}_{\gamma}}.
\label{eq:Ttens}
\ee
The inner product~(\ref{eq:tensdot}) can then be written 
as a left multiplication of the last column-mode of $T$
with the first row-mode of $S$. 
Since $\gamma$ could have been lowered in $S$ (instead of $T$), 
mathematically
this operation is exactly equal to a right multiplication between the
last row-mode of $T$ and the first column-mode of $S$\footnote{
Once the tensor indices are given explicitly, the order of the operands 
does not affect the outcome: 
$ T\indices{^{\alpha \beta}_{\gamma}} 
  S^{\gamma \rho} \equiv S^{\gamma \rho} 
  T\indices{^{\alpha \beta}_{\gamma}} 
  $. 
}
\be
T\indices{^{\alpha \beta}_{\gamma}} S^{\gamma \rho} \equiv
T^{\alpha \beta \gamma} S\indices{_\gamma^\rho}.
\label{eq:Tinner}
\ee
	
The inner product~(\ref{eq:Tinner}) constitutes only one way of
contracting one index of $T$ with one of $S$. In general, depending on
which index is contracted, there are six mathematically distinct
cases:  
\be
\begin{array}[c]{lllll}
  {\rm (i)}\ & 
  T\indices{^{\alpha \beta}_{\sigma}} \; S^{\sigma \rho} = 
  T\indices{^{\alpha \beta \gamma}} \ g_{\gamma \sigma} \ S^{\sigma \rho} 
  & \qquad \qquad & 
  {\rm (ii)}\ & 
  T\indices{^{\alpha \beta}_{\rho}} \; S^{\sigma \rho} = 
  T\indices{^{\alpha \beta \gamma}} \ g_{\gamma \rho} \ S^{\sigma \rho} 
  \\[4pt]
  {\rm (iii)}\ & 
  T\indices{^{\alpha}_{\sigma}^{\gamma}} \; S^{\sigma \rho} = 
  T\indices{^{\alpha \beta \gamma}} \ g_{\beta \sigma} \ S^{\sigma \rho} 
  & \qquad \qquad & 
  {\rm (iv)}\ & 
  T\indices{^{\alpha}_{\rho}^{\gamma}} \; S^{\sigma \rho} = 
  T\indices{^{\alpha \beta \gamma}} \ g_{\beta \rho} \ S^{\sigma \rho} 
  \\[4pt]
  {\rm (v)}\ & 
  T\indices{_{\sigma}^{\beta \gamma}} \; S^{\sigma \rho} = 
  T\indices{^{\alpha \beta \gamma}} \ g_{\alpha \sigma} \ S^{\sigma \rho} 
  & \qquad \qquad & 
  {\rm (vi)}\ & T\indices{_{\rho}^{\beta \gamma}} \; S^{\sigma \rho} = 
  T\indices{^{\alpha \beta \gamma}} \ g_{\alpha \rho} \ S^{\sigma \rho}.
\end{array}
\label{eq:tenscases}
\ee 
Since the index sizes can differ substantially, the distinction
among all these cases is not just formal. Additionally, a tensor can
be symmetric with respect to the interchange of the first and third
index but not with respect to the interchange between the first and
the second. Consequently the mutual ordering among indices is quite
important prompting to distinguish among cases with different
ordering.  Thus not only matters which index is contracted, but also
in which position each index is. Ultimately, a tensor is a numerical
object stored in the physical memory of a computer. We will see that
index structure and storage strategies play a fundamental role in the
optimal use of the BLAS library.

\subsection{Slicing the modes}
\label{sec:slicing}

Here we discuss how the procedure used in Sec.~(\ref{sec:example}) to
slice matrices can be generalized to tensors, to produce
objects manageable by BLAS.  In general, an N-dimensional tensor is
sliced by (N-1)-dimensional objects, and every slice is itself a
tensor. Notice that the slicing language shifts the focus from the
parts of the partitioned tensor to the partitioning object itself.  As
we will see in the next section, this shift in focus is at the core of
our analysis.

\begin{figure}[!htb]
\centering
\begin{tikzpicture}
\draw[white] (0,-2) grid (16,2);
\filldraw[fill=yellow!50] (3,-1) rectangle node[left=30pt]{$\alpha$} node[below=30pt]{$\beta$} node{$T$} (5,1)
	[shift={(4,0)}]  (3,-1) rectangle node[left=30pt]{$\alpha$} node[below=30pt]{$\beta$} node{$T$} (5,1)
	[shift={(4,0)}]  (3,-1) rectangle node[left=30pt]{$\alpha$} node[below=30pt]{$\beta$} node{$T$} (5,1);
\filldraw[fill=yellow!50] (3,1) -- node[above]{$\sigma$} (4,1.5) -- (6,1.5) -- (5,1)
	[shift={(4,0)}] (3,1) -- node[above]{$\sigma$} (4,1.5) -- (6,1.5) -- (5,1)
	[shift={(4,0)}] (3,1) -- node[above]{$\sigma$} (4,1.5) -- (6,1.5) -- (5,1);
\filldraw[fill=yellow!50] (6,1.5) --(6,-0.5) --(5,-1) -- (5,1)
	[shift={(4,0)}] (6,1.5) --(6,-0.5) --(5,-1) -- (5,1)
	[shift={(4,0)}] (6,1.5) --(6,-0.5) --(5,-1) -- (5,1);
\draw[blue,thick] (3.5,-1) -- (3.5,1) -- (4.5,1.5)
	[shift={(8,0)}] (3.5,-1) -- (3.5,1) -- (4.5,1.5);
\draw[red,thick] (7,0.5) -- (9,0.5) -- (10,1)
	[shift={(4,0)}] (7,0.5) -- (9,0.5) -- (10,1); 
\end{tikzpicture}
\caption{Three possible slicing choices}
\label{fig:part1}
\end{figure}
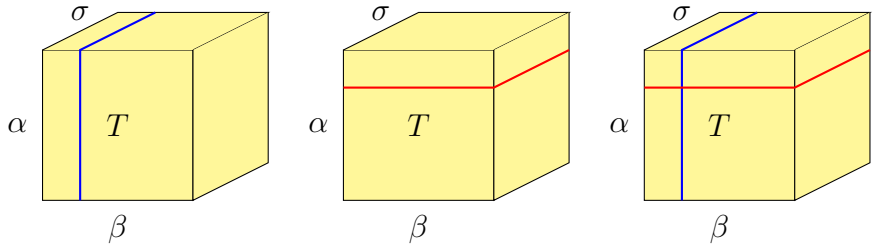

To allow an easy visualization,
let us again consider the contraction case (i) of Eq.~\eqref{eq:tenscases}.
In order to reduce this operation to a sequence of matrix-matrix multiplications,
the (2,1)-tensor $T\indices{^{\alpha\beta}_{\sigma}}$ needs to be sliced 
by a 2-dimensional object; such a slicing can be
realized in several ways. For the sake of simplicity, let us represent 
$T$ as a 3-dimensional cube, and consider its slicing by one or more planes. 
Figure~\ref{fig:part1} shows three (out of seven) possible slicings of 
$T$ by different choices of planes.

\begin{itemize}
\item[(a)] The cube is sliced by a plane defined by the modes $\alpha$
and $\sigma$.
\item[(b)] The cube is sliced by a plane defined by the modes $\beta$ and $\sigma$. 
\item[(c)] This double slicing is defined by combining the two previous slicing, (a) and (b).
\end{itemize}

A simple yet effective way of describing the slicing of a tensor is by means
of a vector \sv\ that indicates the indices that are sliced:
\be
		\vec{s}(T) = (s_\alpha, s_\beta, s_\sigma).
	\ee
In this language, the previous three cases correspond to
\begin{itemize}
	\item[(a)] $\vec{s}_a(T) = (s_\alpha, s_\beta, s_\sigma) = $ (0, 1, 0);
	\item[(b)] $\vec{s}_b(T) =$ (1, 0, 0);
	\item[(c)] $\vec{s}_c (T)=$ (1, 1, 0); 
          this is the sum of the previous two slicings, 
          $\vec{s}_a(T) + \vec{s}_b(T) = \vec{s}_c(T)$.     
\end{itemize}
In the first case, $s_\beta = 1$ means that the index $\beta$ is decomposed 
in a sequence of as many slices as the range of $\beta$. 
Accordingly, $s_\alpha$ and $s_\sigma$ take value 0, i.e., 
the slicing is made by planes defined by the modes $\alpha$ and $\sigma$ along the $\beta$ 
direction. Note that any combination of slicings results in an object of
dimensionality  $N-\|\vec{s}(T)\|_1$. 
For instance, slicing (c) produces a set of $3-2=1$-dimensional objects.   

We now relate the slicing to mappings onto BLAS routines. To this end, 
we have to specify how the tensors are stored in memory. 
In the rest of the paper we always assume the first index
of each tensor to have $stride = 1$.
The same way in Sec.~\ref{sec:example} we decomposed the contraction
between $A$ and $B$, in the following we decompose the contraction 
$R\indices{^{\alpha \beta \rho}} := T\indices{^{\alpha \beta}_{\sigma}} \; S^{\sigma \rho}$
in three different ways.
Here the tensors $T$, $S$ and $R$ are of size $m \times n \times k$, $k
\times \ell$, and $m \times n \times \ell$, respectively. (Then the slices
$R^{\alpha i \rho}$ are of size $m \times \ell$, and 
$T\indices{^{\alpha i}_{\sigma}}$ are of size $m \times k$.)

\begin{itemize}
\item[(a)] $\vec{s}_a(T) = $ (0, 1, 0) leads to a sequence of matrix
  products directly mappable onto \gem s since the dimensions with
  stride=1 remain unsliced: \fbox{$R = \forall_{i} \ T_i S$}.

\begin{tikzpicture}
\draw[white] (-3.5,-2) grid (12,2.5);
\draw (-2.3,-1.28) -- (-3,-1.5) -- node[right=-4pt] {$R^{\alpha 1 \rho}$} (-3,1.5) -- (-3+1,1.8) -- (-2,1.6);
\draw (-1.5,-1.28) -- (-2.2,-1.5) -- node[right=-4pt] {$R^{\alpha 2 \rho}$} (-2.2,1.5) -- (-1.2,1.8) -- (-1.2,1.6);
\draw (-1.4,-1.5) -- node[right=-2pt] {$R^{\alpha 3 \rho}$} (-1.4,1.5) -- (-0.4,1.8) -- (-0.4,-1.2) -- (-1.4,-1.5);
\foreach \x in {0,0.2,0.4,0.6}
	\draw[fill=black] (\x,0) circle(0.05);
\draw (0.8,-1.5) -- node[right=-2pt] {$R^{\alpha n \rho}$} (0.8,1.5) -- (1.8,1.8) -- (1.8,-1.2) -- (0.8,-1.5);
\draw[thick] (2.2,0.07) -- (2.5,0.07);
\draw[thick] (2.2,-0.07) -- (2.5,-0.07);
\draw[thick] (3.1,-1.8) -- (2.7,-1.8) -- (2.7,2) -- (3.1,2);
\draw (2.9+0.7,-1.28) -- (2.9,-1.5) --  node[right=-3pt] {$T\indices{^{\alpha 1}_{\sigma}}$} (2.9,1.5) -- (2.9+1,1.8) -- (2.9+1,1.6);
\draw (3.7+0.7,-1.28) -- (3.7,-1.5) --  node[right=-3pt] {$T\indices{^{\alpha 2}_{\sigma}}$} (3.7,1.5) -- (3.7+1,1.8) -- (3.7+1,1.6);
\draw (4.5,-1.5) -- node[right=-2pt] {$T\indices{^{\alpha 3}_{\sigma}}$} (4.5,1.5) -- (5.5,1.8) -- (5.5,-1.2) -- (4.5,-1.5);
\foreach \x in {5.9,6.1,6.3,6.5}
	\draw[fill=black] (\x,0) circle(0.05);
\draw (6.9,-1.5) -- node[right=-2pt] {$T\indices{^{\alpha n}_{\sigma}}$} (6.9,1.5) -- (7.9,1.8) -- (7.9,-1.2) -- (6.9,-1.5);
\draw[thick] (7.7,-1.8) -- (8.1,-1.8) -- (8.1,2) -- (7.7,2);
\draw[thick] (8.4,-0.2) -- (8.8,0.2);
\draw[thick] (8.4,0.2) --(8.8,-0.2);
\draw (9.3,-1) rectangle node {$S\indices{^{\sigma \rho}}$} (11.3,1);
\end{tikzpicture}
\item[(b)] $\vec{s}_b(T) =$ (1, 0, 0) 
  also leads to a sequence of matrix products; in this case, however, to
  enable the use of \gem, each slice of $T$ must be copied on a separate
  memory region: \fbox{$R = \forall_{i} \ (T_i)_{cp}\ S$}.
 
\begin{tikzpicture}
\draw[white] (1,-2.5) grid (17,2.5);
\draw (3.5,-1.5) -- (3,-2) -- node[above] {$R^{1 \beta \rho}$} (6,-2) -- (7,-1) -- (6.5,-1)
	[shift={(5,0)}] (3.5,-1.5) -- (3,-2) -- node[above=-2pt] {$T\indices{^{1 \beta}_{\sigma}}$} (6,-2) -- (7,-1) -- (6.5,-1)
	[shift={(-5,0.6)}] (3.5,-1.5) -- (3,-2) -- node[above] {$R^{2 \beta \rho}$} (6,-2) -- (7,-1) -- (6.5,-1)
	[shift={(5,0)}] (3.5,-1.5) -- (3,-2) -- node[above=-2pt] {$T\indices{^{2 \beta}_{\sigma}}$} (6,-2) -- (7,-1) -- (6.5,-1);
\draw (3,-0.8) -- node[above] {$R^{3 \beta \rho}$} (6,-0.8) -- (7,0.2) -- (4,0.2) -- (3,-0.8)
	[shift={(5,0)}] (3,-0.8) -- node[above=-2pt] {$T\indices{^{3 \beta}_{\sigma}}$} (6,-0.8) -- (7,0.2) -- (4,0.2) -- (3,-0.8) 
	[shift={(-5,1.9)}] (3,-0.8) -- node[above] {$R^{m \beta \rho}$} (6,-0.8) -- (7,0.2) -- (4,0.2) -- (3,-0.8)
	[shift={(5,0)}] (3,-0.8) -- node[above=-2pt] {$T\indices{^{m \beta}_{\sigma}}$} (6,-0.8) -- (7,0.2) -- (4,0.2) -- (3,-0.8); 
\foreach \y in {-0.1,0.2,0.5,0.8}
	\draw[fill=black] (5,\y) circle(0.05);
\foreach \y in {-0.1,0.2,0.5,0.8}
	\draw[fill=black] (10,\y) circle(0.05);
\draw[thick] (7.5,0.07) -- (7.8,0.07);
\draw[thick] (7.5,-0.07) -- (7.8,-0.07);
\draw[thick] (7.8,-1.8) -- (7.8,-2.2) -- (12.2,-2.2) -- (12.2,-1.8);
\draw[thick] (7.8,1.9) -- (7.8,2.3) -- (12.2,2.3) -- (12.2,1.9);
\draw[thick] (12.6,-0.2) -- (13,0.2);
\draw[thick] (12.6,0.2) --(13,-0.2);
\draw (13.3,-1) rectangle node {$S\indices{^{\sigma \rho}}$} (15.3,1);
\end{tikzpicture}

\item[(c)] $\vec{s}_c(T) =$ (1, 1, 0) leads to a sequence of \gemv s: \fbox{$R = \forall_{i} \forall_{j} \ T_{ij} S$}.\\
\begin{tikzpicture}
\draw[white] (0,-2.5) grid (16,2.5);
\draw[thick,gray] (2,-2) -- (3,-1)
	[shift={(0,0.2)}] (2,-2) -- (3,-1)
	[shift={(0,0.2)}] (2,-2) -- (3,-1)
	[shift={(0,0.2)}] (2,-2) -- (3,-1)
	[shift={(0,0.2)}] (2,-2) -- (3,-1)
	[shift={(0,0.2)}] (2,-2) -- (3,-1)
	[shift={(0,0.2)}] (2,-2) -- (3,-1)
	[shift={(0,0.2)}] (2,-2) node[left=-4pt,black]{\shortstack{$i \uparrow$\\$R^{i 1 \rho}$}} -- (3,-1)
	[shift={(0,0.2)}] (2,-2) -- (3,-1)
	[shift={(0,0.2)}] (2,-2) -- (3,-1)
	[shift={(0,0.2)}] (2,-2) -- (3,-1)
	[shift={(0,0.2)}] (2,-2) -- (3,-1)
	[shift={(0,0.2)}] (2,-2) -- (3,-1)
	[shift={(0,0.2)}] (2,-2) -- (3,-1)
	[shift={(0,0.2)}] (2,-2) -- (3,-1)
	[shift={(0,0.2)}] (2,-2) -- (3,-1);
\draw[thick] (2.3,-2) node[below,right=20pt]{\shortstack{$j \rightarrow$\\$R^{1 j \rho}$}} -- (3.3,-1)
	[shift={(0,0.2)}] (2.3,-2) -- (3.3,-1)
	[shift={(0,0.2)}] (2.3,-2) -- (3.3,-1)
	[shift={(0,0.2)}] (2.3,-2) -- (3.3,-1)
	[shift={(0,0.2)}] (2.3,-2) -- (3.3,-1)
	[shift={(0,0.2)}] (2.3,-2) -- (3.3,-1)
	[shift={(0,0.2)}] (2.3,-2) -- (3.3,-1)
	[shift={(0,0.2)}] (2.3,-2) -- (3.3,-1)
	[shift={(0,0.2)}] (2.3,-2) -- (3.3,-1)
	[shift={(0,0.2)}] (2.3,-2) -- (3.3,-1)
	[shift={(0,0.2)}] (2.3,-2) -- (3.3,-1)
	[shift={(0,0.2)}] (2.3,-2) -- (3.3,-1)
	[shift={(0,0.2)}] (2.3,-2) -- (3.3,-1)
	[shift={(0,0.2)}] (2.3,-2) -- (3.3,-1)
	[shift={(0,0.2)}] (2.3,-2) -- (3.3,-1)
	[shift={(0,0.2)}] (2.3,-2) -- (3.3,-1);
\foreach \x in {3.6,3.9,4.2,4.5}
	\draw[fill=black] (\x,0) circle(0.05);
\draw[thick,gray] (4.7,-2) -- (5.7,-1)
	[shift={(0,0.2)}] (4.7,-2) -- (5.7,-1)
	[shift={(0,0.2)}] (4.7,-2) -- (5.7,-1)
	[shift={(0,0.2)}] (4.7,-2) -- (5.7,-1)
	[shift={(0,0.2)}] (4.7,-2) -- (5.7,-1)
	[shift={(0,0.2)}] (4.7,-2) -- (5.7,-1)
	[shift={(0,0.2)}] (4.7,-2) -- (5.7,-1)
	[shift={(0,0.2)}] (4.7,-2) -- (5.7,-1)
	[shift={(0,0.2)}] (4.7,-2) -- (5.7,-1)
	[shift={(0,0.2)}] (4.7,-2) -- (5.7,-1)
	[shift={(0,0.2)}] (4.7,-2) -- (5.7,-1)
	[shift={(0,0.2)}] (4.7,-2) -- (5.7,-1)
	[shift={(0,0.2)}] (4.7,-2) -- (5.7,-1)
	[shift={(0,0.2)}] (4.7,-2) -- (5.7,-1)
	[shift={(0,0.2)}] (4.7,-2) -- (5.7,-1)
	[shift={(0,0.2)}] (4.7,-2) -- (5.7,-1);
\draw[thick] (5,-2) -- (6,-1)
	[shift={(0,0.2)}] (5,-2) -- (6,-1)
	[shift={(0,0.2)}] (5,-2) -- (6,-1)
	[shift={(0,0.2)}] (5,-2) -- (6,-1)
	[shift={(0,0.2)}] (5,-2) -- (6,-1)
	[shift={(0,0.2)}] (5,-2) -- (6,-1)
	[shift={(0,0.2)}] (5,-2) -- (6,-1)
	[shift={(0,0.2)}] (5,-2) -- (6,-1)
	[shift={(0,0.2)}] (5,-2) -- (6,-1)
	[shift={(0,0.2)}] (5,-2) -- (6,-1)
	[shift={(0,0.2)}] (5,-2) -- (6,-1)
	[shift={(0,0.2)}] (5,-2) -- (6,-1)
	[shift={(0,0.2)}] (5,-2) -- (6,-1)
	[shift={(0,0.2)}] (5,-2) -- (6,-1)
	[shift={(0,0.2)}] (5,-2) -- (6,-1)
	[shift={(0,0.2)}] (5,-2) -- (6,-1);
\draw[thick] (6.5,0.07) -- (6.8,0.07);
\draw[thick] (6.5,-0.07) -- (6.8,-0.07);
\draw[thick] (7.3,2.2) -- (7,2.2) -- (7,0.3);
\draw[thick] (7,-1.1) -- (7,-2.2) -- (7.3,-2.2);
\draw[thick,gray] (7.3,-2) -- (8.3,-1)
	[shift={(0,0.2)}] (7.3,-2) -- (8.3,-1)
	[shift={(0,0.2)}] (7.3,-2) -- (8.3,-1)
	[shift={(0,0.2)}] (7.3,-2) -- (8.3,-1)
	[shift={(0,0.2)}] (7.3,-2) -- (8.3,-1)
	[shift={(0,0.2)}] (7.3,-2) -- (8.3,-1)
	[shift={(0,0.2)}] (7.3,-2) -- (8.3,-1)
	[shift={(0,0.2)}] (7.3,-2) node[left=-6pt,black]{\shortstack{$i \uparrow$\\$T\indices{^{i 1}_{\sigma}}$}} -- (8.3,-1)
	[shift={(0,0.2)}] (7.3,-2) -- (8.3,-1)
	[shift={(0,0.2)}] (7.3,-2) -- (8.3,-1)
	[shift={(0,0.2)}] (7.3,-2) -- (8.3,-1)
	[shift={(0,0.2)}] (7.3,-2) -- (8.3,-1)
	[shift={(0,0.2)}] (7.3,-2) -- (8.3,-1)
	[shift={(0,0.2)}] (7.3,-2) -- (8.3,-1)
	[shift={(0,0.2)}] (7.3,-2) -- (8.3,-1)
	[shift={(0,0.2)}] (7.3,-2) -- (8.3,-1);
\draw[thick] (7.6,-2) node[below,right=20pt]{\shortstack{$j \rightarrow$\\$T\indices{^{1 j}_{\sigma}}$}} -- (8.6,-1)
	[shift={(0,0.2)}] (7.6,-2) -- (8.6,-1)
	[shift={(0,0.2)}] (7.6,-2) -- (8.6,-1)
	[shift={(0,0.2)}] (7.6,-2) -- (8.6,-1)
	[shift={(0,0.2)}] (7.6,-2) -- (8.6,-1)
	[shift={(0,0.2)}] (7.6,-2) -- (8.6,-1)
	[shift={(0,0.2)}] (7.6,-2) -- (8.6,-1)
	[shift={(0,0.2)}] (7.6,-2) -- (8.6,-1)
	[shift={(0,0.2)}] (7.6,-2) -- (8.6,-1)
	[shift={(0,0.2)}] (7.6,-2) -- (8.6,-1)
	[shift={(0,0.2)}] (7.6,-2) -- (8.6,-1)
	[shift={(0,0.2)}] (7.6,-2) -- (8.6,-1)
	[shift={(0,0.2)}] (7.6,-2) -- (8.6,-1)
	[shift={(0,0.2)}] (7.6,-2) -- (8.6,-1)
	[shift={(0,0.2)}] (7.6,-2) -- (8.6,-1)
	[shift={(0,0.2)}] (7.6,-2) -- (8.6,-1);
\foreach \x in {8.9,9.2,9.5,9.8}
	\draw[fill=black] (\x,0) circle(0.05);
\draw[thick,gray] (10.3,-2) -- (11.3,-1)
	[shift={(0,0.2)}] (10.3,-2) -- (11.3,-1)
	[shift={(0,0.2)}] (10.3,-2) -- (11.3,-1)
	[shift={(0,0.2)}] (10.3,-2) -- (11.3,-1)
	[shift={(0,0.2)}] (10.3,-2) -- (11.3,-1)
	[shift={(0,0.2)}] (10.3,-2) -- (11.3,-1)
	[shift={(0,0.2)}] (10.3,-2) -- (11.3,-1)
	[shift={(0,0.2)}] (10.3,-2) -- (11.3,-1)
	[shift={(0,0.2)}] (10.3,-2) -- (11.3,-1)
	[shift={(0,0.2)}] (10.3,-2) -- (11.3,-1)
	[shift={(0,0.2)}] (10.3,-2) -- (11.3,-1)
	[shift={(0,0.2)}] (10.3,-2) -- (11.3,-1)
	[shift={(0,0.2)}] (10.3,-2) -- (11.3,-1)
	[shift={(0,0.2)}] (10.3,-2) -- (11.3,-1)
	[shift={(0,0.2)}] (10.3,-2) -- (11.3,-1)
	[shift={(0,0.2)}] (10.3,-2) -- (11.3,-1);
\draw[thick] (10.6,-2) -- (11.6,-1)
	[shift={(0,0.2)}] (10.6,-2) -- (11.6,-1)
	[shift={(0,0.2)}] (10.6,-2) -- (11.6,-1)
	[shift={(0,0.2)}] (10.6,-2) -- (11.6,-1)
	[shift={(0,0.2)}] (10.6,-2) -- (11.6,-1)
	[shift={(0,0.2)}] (10.6,-2) -- (11.6,-1)
	[shift={(0,0.2)}] (10.6,-2) -- (11.6,-1)
	[shift={(0,0.2)}] (10.6,-2) -- (11.6,-1)
	[shift={(0,0.2)}] (10.6,-2) -- (11.6,-1)
	[shift={(0,0.2)}] (10.6,-2) -- (11.6,-1)
	[shift={(0,0.2)}] (10.6,-2) -- (11.6,-1)
	[shift={(0,0.2)}] (10.6,-2) -- (11.6,-1)
	[shift={(0,0.2)}] (10.6,-2) -- (11.6,-1)
	[shift={(0,0.2)}] (10.6,-2) -- (11.6,-1)
	[shift={(0,0.2)}] (10.6,-2) -- (11.6,-1)
	[shift={(0,0.2)}] (10.6,-2) -- (11.6,-1);
\draw[thick] (11.7,2.2) -- (12,2.2) -- (12,-2.2) -- (11.7,-2.2);
\draw[thick] (12.6,-0.2) -- (13,0.2);
\draw[thick] (12.6,0.2) --(13,-0.2);
\draw (13.3,-1) rectangle node {$S\indices{^{\sigma \rho}}$} (15.3,1);
\end{tikzpicture}
\end{itemize}

Once again, these three slicings
lead to mathematically equivalent algorithms for computing the tensor $R$. In
terms of performance, (a) is to be preferred, as it directly maps onto matrix
products. Although slicing (b) also reduces the problem to a sequence of
matrix-matrix products, 
since the slices have both dimensions with stride $>1$,
these cannot be directly mapped onto \gem; this means that a costly preliminary
transposition is necessary. Finally, slicing (c) should be avoided, as it
forms $R$ by means of matrix-vector products, thus relying on the slow BLAS 2 kernel.

\section{The recipe for an efficient use of BLAS}
\label{sec:sec3}
Given two tensors of arbitrary dimension and one or more indices to be
contracted, for performance reasons the objective is to map the
calculations onto BLAS 3 kernels, and onto \gem\ in particular.  The
mapping is obtained by slicing the tensors opportunely.
Unfortunately, not every contraction can be expressed in terms of
level 3 kernels, and even when that is possible, not every choice of
slicing leads to the direct use of \gem. In this section, we first
show that by testing a small set of conditions it is possible to
determine when and how \gem\ can be employed.  Then, according to
these conditions, we classify the contractions in three groups, and
for each group we provide a recipe on how to best utilize BLAS.

Ultimately, our tools are meant to help application scientists 
in two different scenarios. If they have direct control of 
the generation of the tensors, we indicate a clear strategy to store them 
so that the contractions can be performed avoiding costly memory operations.
Conversely, if the storage scheme of the tensors is fixed, we give
guidelines to use the full potential of BLAS.

\subsection{Requirements to be satisfied for a BLAS 3 mapping}
\label{sec:req}
As illustrated in Sec.~\ref{sec:slicing}, a slicing transforms a
contraction between two generic tensors $T$ and $S$ into a sequence
(or a sum) of contractions between the slices of $T$ and those of $S$.
To allow a mapping onto BLAS 3, both the slices of $S$ and $T$ need to
conform to the \gem\ interface (see Sec.~\ref{sec:blasdef}).
In the first place, they have to be matrices; 
hence $S$ and $T$ have to have at least
$N=2$ indices, and if $N > 2$, the tensor has to be sliced along $N-2$
of them. Moreover, the slices of $S$ and $T$ must
have exactly one contracted and one free (non-contracted) index; in
other words, we are looking for two-dimensional slices that share one
dimension, the same way in the contraction $C:= A B$, the matrices
$A\in \mathbf{R}^{m\times k}$ and $B\in \mathbf{R}^{k\times n}$ share
the $k$-dimension, while $m$ and $n$ are free. Finally, the BLAS
interface does not allow the use of matrices with stride $> 1$ in both
dimensions; therefore it is not possible to slice the indices of $S$
and $T$ that correspond to stride 1 modes without compromising the use
of \gem.

The previous considerations can be formalized as three requirements.
Define $N(X)$ as the number of modes of tensor $X$, 
and let $R := T S$ be a contraction.  
\begin{itemize}
\item[R1.] The modes of $T$ and $S$ with stride 1 must not be sliced.
\item[R2.] $T$ and $S$ have to be sliced along $N(T)-2$ and $N(S)-2$ modes, respectively.
  (In a compact language,  $\|\vec{s}(T)\|_1 = N(T)-2$ and $\|\vec{s}(S)\|_1 = N(S)-2$).
  Notice that whenever a contracted index is sliced, 
  both $T$ and $S$ are affected.
\item[R3.] 
  The slices must have exactly one free and one contracted index.
  Therefore, it must be avoided to slice only all the contracted 
  indices or only all the free ones.
\end{itemize}

Despite their simplicity, these three requirements 
apply universally to contractions $R = TS$ of any order between tensors of 
generic dimensionality. 
If they are all satisfied, then it is possible to compute $R$ via \gem s;
this is for instance the case of Example (a) in Sec.~\ref{sec:slicing}.

A precise classification of contractions is given in the next
section. Let us now briefly discuss the scenarios in which exactly one of the
requirements does not hold.
\bi
\item[F1.] If requirement R1 is not satisfied (the resulting slice
  does not have any mode with stride 1), the use of \gem\ is still
  possible by copying each slice in a separate portion of memory.  We
  refer to this operation as a \copgem , an example of whom is given
  by case (b) in Sec.~\ref{sec:slicing}.

\item[F2.] If $\|\vec{s}(T)\|_1 > N(T)-2$ or $\|\vec{s}(S)\|_1 > N(S)-2$, 
then it is not possible to use BLAS 3 subroutines; 
in this case, one has to rely on routines from level 2 or level 1, such as
\gemv\ and \ger, or even on no BLAS at all. 
An example is provided by case (c) in Sec.~\ref{sec:slicing}.

\item[F3.]  As for F2, if all the contracted indices or all the free
  indices are sliced, one can use only BLAS 2, BLAS 1, or even no BLAS
  at all.  An example is given by the contraction $R\indices{^{\alpha
      \beta \rho}} := T\indices{^{\alpha \beta}_{\sigma}} \; S^{\sigma
    \rho}$ with $\vec{s}(T)=(0,0,1)$: the computation reduces to a
  sequence of element-wise operations.
\ei

\subsection{Contraction classes}
\label{sec:classes}
We now 
use the requirements for cataloging all contractions in a
small set of classes; most importantly, for each class we provide a
simple recipe, indicating how to achieve an optimal use of BLAS. The
aim is to empower computational scientists with a precise and 
easy-to-implement tool: Ultimately, these recipes shed light on the 
importance of choosing the correct strategy for generating and 
storing tensors in memory.

Given two tensors $t_1$ and $t_2$, and the number of contractions $p$,
let us define $\Delta(t_i)$ as $N(t_i)-p$, with $i = 1, 2$
($\Delta$ counts the number of indices of $t_i$ minus the
number of contracted indices, that is, the number of free indices). 
The following classification 
is entirely based on $\Delta(t_1)$ and $\Delta(t_2)$.

\subsubsection{Class 1: $\Delta(t_1)=0$ and $\Delta(t_2)=0$} 
The tensors have the same number of modes and they are all contracted;
this operation produces a scalar. For example, 
\begin{equation}
\kappa := T\indices{^{\alpha\beta\gamma}} S\indices{_{\alpha\beta\gamma}}.
\label{eq:class1}
\end{equation}
This situation is rather uncommon, yet possible. 
It can be considered as an extreme example of F3 in which there are no free
indices; \gem\ is therefore not applicable.  
Moreover, since BLAS does not support operations with more than one 
contracted index, one is forced to slice $k-1$ indices, 
resulting in an element-wise product and a sum-reduction to
a scalar. The only opportunity for BLAS lies in level 1.  
For instance, in Eq.~\eqref{eq:class1} it is possible to slice the indices 
$\alpha$ and $\beta$, $\vec{s}(T)=(1,1,0) \equiv \vec{s}(S)=(1,1,0)$; in both
tensors, the slices are thus vectors along $\gamma$, and participate in 
\dt{} operations.

\paragraph{RECIPE, Class 1.}
{\em Slice all but one index. Use \dt{} with the unsliced mode. 
Then sum-reduce the result of all \dt{}'s.}

\subsubsection{Class 2: $\Delta(t_1)\geq1$ and $\Delta(t_2)=0$}
This class contains all the contractions in which only one tensor 
has one or more free indices.
Let us look for example at
$$
R\indices{^{\mu_1 \dots \mu_n}} := 
T\indices{^{\mu_1 \dots \mu_n \beta \gamma}} S_{\beta \gamma}.$$
Requirement R2 forces $T$ to be sliced $n$ times. 
Moreover, since BLAS allows at most one contracted index, 
$S$ still requires to be sliced along $N(S)-1$ indices,
leading to a sequence of vectors. 
This means that there is no possible choice that allows the use of \gem: 
At best, the contraction is computed via matrix-vector operations. 

Motivated by Eq.~(\ref{eq:tenscases}),
we use the next three double contractions as explanatory examples.
\be
(1')\ R^{\alpha} := T\indices{^{\alpha \beta \gamma}} g_{\beta \gamma} 
\qquad 
(2')\ R^{\gamma} := T\indices{^{\alpha \beta \gamma}} g_{\alpha \beta} 
\qquad 
(3')\ R^{\beta} := T\indices{^{\alpha \beta \gamma}} g_{\alpha \gamma}.
\label{eq:expls1}
\ee
In order to make use of BLAS routines, 
$T$ has to be sliced at least once. 
For the contraction ($1'$), there are three possible choices:
\bi
\item $\vec{s}(T)=(0,0,1)$, and consequently $\vec{s}(g)=(0,1)$. 
  This option does not slice the stride 1 modes but falls under the case F2, 
  thus only \gemv\ is available;
\item $\vec{s}(T)=(0,1,0)$, and consequently $\vec{s}(g)=(1,0)$. 
  This situation is completely analogous to the previous one. 
 \item $\vec{s}(T)=(1,0,0)$ and $\vec{s}(g)=(0,0)$.  
  Each slice of $T$ has all its indices contracted with those of $g$. 
  Although the index with stride 1 is sliced (case F1),
  this example falls under the case F3, 
  where we sliced the only free index of $T$.
  As a consequence, the use of BLAS is limited to \dt.
 \ei

The analysis of contractions ($2'$) and ($3'$) goes along the same lines:
both cases present a choice of slicing leading to a \gemv\ and a \dt . 
The only additional situation arises when $\vec{s}(T)=(1,0,0)$, 
and consequently $\vec{s}(g)=(1,0)$; 
this choice slices the index of $T$ which carries stride 1. 
Thus, it is not possible to use \gemv, unless each slice of $T$ is 
transposed; we refer to this operation as \copgemv.\\
		
\paragraph{RECIPE, Class 2,}
{\em 
If the index of $t_1$ with stride 1 is contracted, then 
slice all the other contracted indices, and
all the free indices but one.
Vice versa, 
if the index of $t_1$ with stride 1 is free, then 
slice all the other free indices, and
all the contracted indices but one.
In all cases, the use of \gemv\ is possible. 
As a rule of thumb to maximize performance, 
keep the indices with largest size unsliced.
}

\subsubsection{Class 3: $\Delta(t_1)\geq1$ and $\Delta(t_2)\geq1$}
\label{sec:class3}
From a computational point of view, 
this is the most interesting class of contractions.
Both tensors have free indices, and a mapping onto 
either \gem{} or \copgem\ always exists.
In fact, according to requirement R2, 
both tensors have two or more indices and can
therefore be sliced into a series of matrices.  
Moreover, both tensors have at least a
free index and at least a contracted index, also satisfying requirement R3.

To express these contractions as matrix-matrix products---but not
necessarily as straight \gem s---it suffices to slice the tensors so to create 
matrices with a single (shared) contracted index, 
and a free one. 
Whether requirement R1 is satisfied too---thus being able to use \gem--- 
or not, splits this class in two subclasses.

\noindent
{\bf Class 3.1}: 
The indices with stride 1 of $t_1$ and $t_2$ are distinct, and both are contracted.
In order to satisfy requirement R3 
(creating slices with one free and one contracted index),
only one of these indices can be left untouched, 
while the other has to be sliced.
This leads to two alternatives: 
a) proceed to use \copgem{}, and 
b) map onto BLAS 2 kernels.

\noindent
{\bf Class 3.2}: 
The indices with stride 1 of $t_1$ and $t_2$ are the same, 
or are distinct and only one of them is contracted.
There is enough freedom to always avoid slicing the indices of stride 1 
in both tensors, thus granting viable map(s) to \gem.

To illustrate the differences between these two sub-classes, 
we discuss, as a test case, the double contractions between two 3-dimensional
tensors, $A^{\alpha \beta \gamma}$ and $B_{\theta \eta \rho}$. 
In general, ignoring internal symmetries of the operands and assuming that all dimensions match,
$A$ and $B$ can be contracted in 18 distinct ways.
In Table~\ref{tab:new3D} we divide the contractions in three groups:
while the positions and the labels of the indices of $A$ are fixed, 
each group has a different combination of contracted indices with $B$. 
As for the previous examples, we assume that each operand is stored in 
memory so that its first index has stride 1.
\begin{table}[tb]
\caption{Double contractions between two 3-dimensional tensors}
\begin{center}
\begin{tabular}{c |c c|c c|c c|}
\multicolumn{1}{c}{}
& \multicolumn{6}{c}{$\Delta(A) = 1$ and $\Delta(B) = 1$} 
\\ \cline{2-7}

Free index in $B$& 
\multicolumn{2}{|c|}{$\beta$ and $\gamma$ contracted} & 
\multicolumn{2}{c|}{$\alpha$ and $\gamma$ contracted} & 
\multicolumn{2}{c|}{$\alpha$ and $\beta$ contracted} 
\\ \hline
3rd & $A^{\alpha \beta \gamma} B_{\beta \gamma \eta}$ & $A^{\alpha \beta  \gamma} B_{\gamma \beta \eta}$ 
    & $A^{\alpha \beta \gamma} B_{\alpha \gamma \eta}$ & 
      \textcolor{red}{$\mathbf{A^{\alpha \beta \gamma} B_{\gamma \alpha \eta}}$} 
    &  $A^{\alpha \beta \gamma} B_{\alpha \beta \eta}$ & $A^{\alpha \beta \gamma} B_{\beta \alpha \eta}$ 
\\ 
2nd & $A^{\alpha \beta \gamma} B_{\beta \eta \gamma}$ & $A^{\alpha \beta
  \gamma} B_{\gamma \eta \beta}$ & $A^{\alpha \beta \gamma} B_{\alpha \eta
  \gamma}$ &  $A^{\alpha \beta \gamma} B_{\gamma \eta \alpha}$ &  $A^{\alpha
  \beta \gamma} B_{\alpha \eta \beta}$ &  $A^{\alpha \beta \gamma} B_{\beta
  \eta \alpha}$ 
\\ 
1st & \textcolor{red}{$\mathbf{A^{\alpha \beta \gamma} B_{\eta \beta
      \gamma}}$} & $A^{\alpha \beta \gamma} B_{\eta \gamma \beta}$ &
$A^{\alpha \beta \gamma} B_{\eta \alpha \gamma}$ &  $A^{\alpha \beta \gamma}
B_{\eta \gamma \alpha}$ &  $A^{\alpha \beta \gamma} B_{\eta \alpha \beta}$ &
$A^{\alpha \beta \gamma} B_{\eta \beta \alpha}$ \\ 
\hline
\end{tabular}
\end{center}
\label{tab:new3D}
\end{table}
We analyze the two cases highlighted in the table; 
it will then become clear that the remaining
16 contractions belong to either Class 3.1 (12 cases) or Class 3.2 (4 cases).
These same two cases will serve in the next section 
as a template for the numerical tests.

\begin{figure}[!bt]
\centering
\begin{tikzpicture}[scale=0.6]
\draw[white] (0,-10) grid (28,2);
\filldraw[fill=yellow!50] (1,-1) rectangle node{$R$} node[left=15pt]{$\beta$} node[below=15pt]{$\eta$} (3,1)
	[shift={(14,0)}] (1,-1) rectangle  node{$R$} node[left=15pt]{$\beta$} node[below=15pt]{$\eta$} (3,1)
	[shift={(-7,-6)}] (1,-1) rectangle node{$R$} node[left=15pt]{$\beta$} node[below=15pt]{$\eta$} (3,1);
\filldraw[fill=yellow!50] (5,-1) rectangle node[left=15pt]{$\alpha$} node[below=15pt]{$\beta$} node{$A$} (7,1)
	[shift={(5,0)}]  (5,-1) rectangle node[left=15pt]{$\gamma$} node[below=15pt]{$\alpha$} node{$B$} (7,1)
	[shift={(9,0)}]  (5,-1) rectangle node[left=15pt]{$\alpha$} node[below=15pt]{$\beta$} node{$A$} (7,1)
	[shift={(5,0)}]  (5,-1) rectangle node[left=15pt]{$\gamma$} node[below=15pt]{$\alpha$} node{$B$} (7,1)
	[shift={(-12,-6)}]  (5,-1) rectangle node[left=15pt]{$\alpha$} node[below=15pt]{$\beta$} node{$A$} (7,1)
	[shift={(5,0)}]  (5,-1) rectangle node[left=15pt]{$\gamma$} node[below=15pt]{$\alpha$} node{$B$} (7,1);
\filldraw[fill=yellow!50] (5,1) -- node[above=-2pt]{$\gamma$} (6,1.5) -- (8,1.5) -- (7,1)
	[shift={(5,0)}] (5,1) -- node[above=-2pt]{$\eta$} (6,1.5) -- (8,1.5) -- (7,1)
	[shift={(9,0)}] (5,1) -- node[above=-2pt]{$\gamma$} (6,1.5) -- (8,1.5) -- (7,1)
	[shift={(5,0)}] (5,1) -- node[above=-2pt]{$\eta$} (6,1.5) -- (8,1.5) -- (7,1)
	[shift={(-12,-6)}] (5,1) -- node[above=-2pt]{$\gamma$} (6,1.5) -- (8,1.5) -- (7,1)
	[shift={(5,0)}] (5,1) -- node[above=-2pt]{$\eta$} (6,1.5) -- (8,1.5) -- (7,1);
\filldraw[fill=yellow!50] (8,1.5) --(8,-0.5) -- node[below=27pt]{(a) Slicing $\gamma \rightarrow$ \copgem} (7,-1) -- (7,1)
	[shift={(5,0)}] (8,1.5) --(8,-0.5) -- (7,-1) -- (7,1)
	[shift={(9,0)}] (8,1.5) --(8,-0.5) -- node[below=27pt]{(b) Slicing $\gamma$ and $\eta \rightarrow$ \gemv} (7,-1) -- (7,1)
	[shift={(5,0)}] (8,1.5) --(8,-0.5) -- (7,-1) -- (7,1)
	[shift={(-12,-6)}] (8,1.5) --(8,-0.5) -- node[below=27pt]{(c) Slicing $\gamma$ and $\alpha \rightarrow$ \ger} (7,-1) -- (7,1)
	[shift={(5,0)}] (8,1.5) --(8,-0.5) -- (7,-1) -- (7,1);
\draw[blue,thick] (10,0.5) -- (12,0.5) -- (13,1)
	[shift={(14,0)}] (10,0.5) -- (12,0.5) -- (13,1)
	[shift={(-12,-6)}] (10,0.5) -- (12,0.5) -- (13,1)
	[shift={(5,0)}] (10,0.5) -- (12,0.5) -- (13,1);
\draw[green,thick] (5.4,1.2) -- (7.4,1.2) -- (7.4,-0.8)
	[shift={(14,0)}] (5.4,1.2) -- (7.4,1.2) -- (7.4,-0.8)
	[shift={(5,0)}] (5.4,1.2) -- (7.4,1.2) -- (7.4,-0.8)
	[shift={(-12,-6)}] (5.4,1.2) -- (7.4,1.2) -- (7.4,-0.8);
\draw[green,thick] (15.6,-1) -- (15.6,1);
\draw[red,thick] (17.5,-7) -- (17.5,-5) -- (18.5,-4.5);
\draw[black,thick] (3.5,-0.1) -- (4,-0.1)
	[shift={(14,0)}] (3.5,-0.1) -- (4,-0.1)
	[shift={(-7,-6)}] (3.5,-0.1) -- (4,-0.1);
\draw[black,thick] (3.5,0.1) -- (4,0.1)
	[shift={(14,0)}] (3.5,0.1) -- (4,0.1)
	[shift={(-7,-6)}] (3.5,0.1) -- (4,0.1);
\draw[black,thick] (9,0.2) -- (8.6,-0.2)
	[shift={(14,0)}] (9,0.2) -- (8.6,-0.2)
	[shift={(-7,-6)}] (9,0.2) -- (8.6,-0.2);
\draw[black,thick] (9,-0.2) -- (8.6,0.2)
	[shift={(14,0)}] (9,-0.2) -- (8.6,0.2)
	[shift={(-7,-6)}] (9,-0.2) -- (8.6,0.2);
\end{tikzpicture}
\caption{ 
  $R\indices{^\beta_\eta} := A^{\alpha \beta \gamma} B_{\gamma \alpha \eta}$: 
  The direct use of \gem\ is not possible;
  different slicing choices lead to different kernels.
}
\label{fig:3D-1}
\end{figure}

The first contraction, 
$R\indices{^\beta_\eta}:= A^{\alpha \beta \gamma} B_{\gamma \alpha \eta}$,  
belongs to Class 3.1;
no choice of slicing allows the direct use of \gem. 
By slicing the first index of one of the two tensors, 
i.e., either $\vec{s}(A) = (1,0,0)$ (and $\vec{s}(B) = (0,1,0)$), 
or as in Fig.~\ref{fig:3D-1}(a), 
$\vec{s}(B) = (1,0,0)$ (and $\vec{s}(A) = (0,0,1)$),
one ends up with \copgem s.
Alternatively, either one or both tensors can be sliced twice; 
Figures~\ref{fig:3D-1}(b) and (c) diplays two possible choices:
the first one---$\vec{s}_A = (0,0,1)$ and 
$\vec{s}_B = (1,0,1)$---violates requirement R2 and leads to \gemv s, 
while the second one---$\vec{s}_A = (1,0,1)$ and $\vec{s}_B =
(1,1,0)$---violates requirements R2 and R3, and produces \ger s.

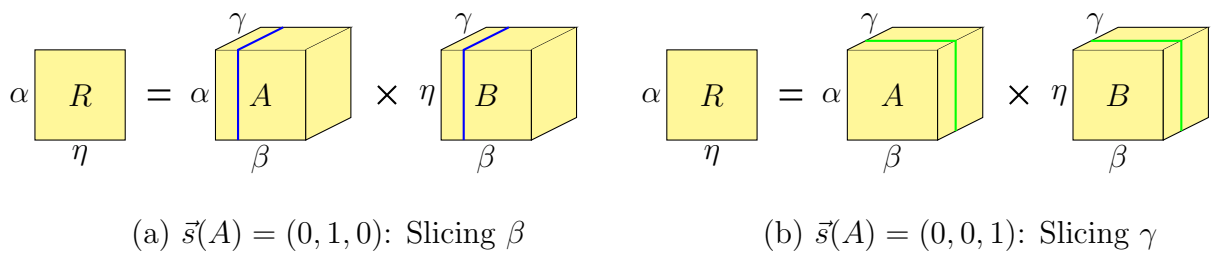
\begin{figure}[!tb]
\centering
\begin{tikzpicture}[scale=0.6]
\draw[white] (0,-4) grid (28,2);
\filldraw[fill=yellow!50] (1,-1) rectangle node{$R$} node[left=15pt]{$\alpha$} node[below=15pt]{$\eta$} (3,1)
	[shift={(14,0)}] (1,-1) rectangle node{$R$} node[left=15pt]{$\alpha$} node[below=15pt]{$\eta$} (3,1);
\filldraw[fill=yellow!50] (5,-1) rectangle node[left=15pt]{$\alpha$} node[below=15pt]{$\beta$} node{$A$} (7,1)
	[shift={(5,0)}]  (5,-1) rectangle node[left=15pt]{$\eta$} node[below=15pt]{$\beta$} node{$B$} (7,1)
	[shift={(9,0)}]  (5,-1) rectangle node[left=15pt]{$\alpha$} node[below=15pt]{$\beta$} node{$A$} (7,1)
	[shift={(5,0)}]  (5,-1) rectangle node[left=15pt]{$\eta$} node[below=15pt]{$\beta$} node{$B$} (7,1);
\filldraw[fill=yellow!50] (5,1) -- node[above=-2pt]{$\gamma$} (6,1.5) -- (8,1.5) -- (7,1)
	[shift={(5,0)}] (5,1) -- node[above=-2pt]{$\gamma$} (6,1.5) -- (8,1.5) -- (7,1)
	[shift={(9,0)}] (5,1) -- node[above=-2pt]{$\gamma$} (6,1.5) -- (8,1.5) -- (7,1)
	[shift={(5,0)}] (5,1) -- node[above=-2pt]{$\gamma$} (6,1.5) -- (8,1.5) -- (7,1);
\filldraw[fill=yellow!50] (8,1.5) --(8,-0.5) -- node[below=30pt]{(a) $\vec{s}(A) = (0,1,0)$: Slicing $\beta$} (7,-1) -- (7,1)
	[shift={(5,0)}] (8,1.5) --(8,-0.5) -- (7,-1) -- (7,1)
	[shift={(9,0)}] (8,1.5) --(8,-0.5) -- node[below=30pt]{(b) $\vec{s}(A) = (0,0,1)$: Slicing $\gamma$} (7,-1) -- (7,1)
	[shift={(5,0)}] (8,1.5) --(8,-0.5) -- (7,-1) -- (7,1);
\draw[blue,thick] (5.5,-1) -- (5.5,1) -- (6.5,1.5)
	[shift={(5,0)}] (5.5,-1) -- (5.5,1) -- (6.5,1.5);
\draw[green,thick] (19.4,1.2) -- (21.4,1.2) -- (21.4,-0.8)
	[shift={(5,0)}] (19.4,1.2) -- (21.4,1.2) -- (21.4,-0.8);
\draw[black,thick] (3.5,-0.1) -- (4,-0.1)
	[shift={(14,0)}] (3.5,-0.1) -- (4,-0.1);
\draw[black,thick] (3.5,0.1) -- (4,0.1)
	[shift={(14,0)}] (3.5,0.1) -- (4,0.1);
\draw[black,thick] (9,0.2) -- (8.6,-0.2)
	[shift={(14,0)}] (9,0.2) -- (8.6,-0.2);
\draw[black,thick] (9,-0.2) -- (8.6,0.2)
	[shift={(14,0)}] (9,-0.2) -- (8.6,0.2);
\end{tikzpicture}
\caption{Two slicing choices for 
  $R\indices{^\alpha_\eta} := A^{\alpha \beta \gamma} B_{\eta \beta \gamma}$, 
  both leading to \gem. }
\label{fig:3D-2}
\end{figure}

The second case, 
$R\indices{^\beta_\eta} := A^{\alpha \beta \gamma} B_{\eta \beta \gamma}$, 
belongs to class 3.2, and is an example in which multiple choices of
slicing equivalently result in \gem.
In fact, 
both $\vec{s}(A) = (0,1,0)$  (and consequently $\vec{s}(B) = (0,1,0)$), 
and $\vec{s}(A) = (0,0,1)$ (and consequently $\vec{s}(B) = (0,0,1)$)
are valid slicings that lead to \gem\ (see Fig.~\ref{fig:3D-2}).

For all the remaining 16 cases of Table~\ref{tab:new3D},
the same considerations hold:
If $A$ and $B$ are stored so that their first indices are either not
contracted, or contracted and the same, then \gem\ can be used directly.
Otherwise, one must pay the overhead due to copying of data (\copgem), or
rely on less efficient lower level BLAS.

Finally, inspired by the contractions arising frequently in   
Quantum Chemistry and in General Relativity, 
we cover now one additional example that involves two tensors of order four.
Instead of describing all the possible cases, we only present two 
double contractions between a (4,0)-tensor $X$ and a (2,2)-tensor $Y$,
again corresponding to classes 3.1 and 3.2.
\be
\label{eq:xy}
	(a)\quad R^{\alpha \beta \gamma \delta} := 
        X^{i \alpha j \beta} Y\indices{_i^\gamma_j^\delta} 
        \qquad\qquad\qquad 
        (b)\quad R^{\alpha \beta \gamma \delta} := 
        X^{i \alpha j \beta} Y\indices{_j^\gamma_i^\delta}.
\ee
These contractions only differ for the relative position of the indices $i$
and $j$ in the tensor $Y$.\footnote{Although confusing, the use of a mixture
  of greek and latin indices will serve a particular purpose when the same
  operations will be considered in the numerical experiments of section
  \ref{sec:appl}}

Case ($a$) admits more than one choice to perform \gem. As an
example, the slicing vectors $\vec{s}(X)=(0,0,1,1)$ and
$\vec{s}(Y)=(0,0,1,1)$ satisfy all three requirements R1, R2, and R3. 
In case ($b$), due to the fact that both contracted indices $i$ and $j$ are the indices
with stride 1, \gem\ cannot be used directly. 
Here we list three possibilities, each one originating computation that
matches a different BLAS routine.
\bi
\item $\vec{s}(X)=(0,0,1,1)$ and $\vec{s}(Y)=(1,1,0,0)$ $\longrightarrow$ \copgem ; \\
  In this case, we violate R1. 
  Since the first index of $Y$ is sliced, the resulting matrices (identified 
  by the values of $j$ and $\gamma$) do not have any dimension with
  stride 1.
  Consequently each slice has to be copied on a separate memory
  allocation before \gem\ can be used.
\item $\vec{s}(X)=(0,1,0,1)$ and $\vec{s}(Y)=(0,1,0,1)$ $\longrightarrow$ \dt ; \\
  Instead of R1, here we violate R3. 
  All the free indices of both tensors are sliced, 
  forcing the double contraction to be computed simultaneously: 
  This is equivalent to compute a series of inner products and sum over them.  
\item $\vec{s}(X)=(1,0,1,1)$ and $\vec{s}(Y)=(1,0,1,1)$ $\longrightarrow$ \ger .\\
	Both R2 and R3 are violated. 
  All contracted indices are sliced, 
  and only two free indices (one for each tensor) are left untouched. 
  This corresponds to a series of outer products to be summed over.
\ei
 
In Sec.~\ref{sec:appl}, we use exactly these slicing choices 
to give an idea of the performance of these different routines in
the case of (highly skewed) rectangular tensors.

\paragraph{RECIPE, Class 3.1.}
{\em 
The objective is to use \copgem{}.
Therefore, for either $t_1$ or $t_2$, keep the first index untouched,
and slice all the others contracted indices; 
for the other, slice all the contracted indices but one. 
In both tensors, slice all the free indices but one. 
In general, the best performance is attained when the dimension of largest
size is left untouched; however, the actual efficiency depends on the 
actual BLAS implementation.
If possible, avoid this class altogether by storing the tensor in memory so
that the contraction falls in class 3.2
}

\paragraph{RECIPE, Class 3.2.}
{\em 
\gem\ is always possible. 
Leave both the stride-1 indices untouched. 
Slice the other indices to satisfy requirements R2 and R3.
For best performance, keep untouched the index with largest size.
}\\

\subsection{BLAS and memory storage}
\label{sec:stor}

Independently of the order of the tensors $t_1$ and $t_2$ and of the number of
contractions, our classification by means of $\Delta(t_i)$ is a general tool
to easily determine whether \gem\ is usable or not.  For a given operation,
the user can immediately determine the class to which it belongs. If both
$\Delta(t_1)$ and $\Delta(t_2)$ equal 0, it is only possible to use \dt. If
exactly one between $\Delta(t_1)$ and $\Delta(t_1)$ equals 0, then BLAS 2
operations are usable. Finally, the most interesting scenario is when both
$\Delta(t_1)$ and $\Delta(t_2)$ are greater or equal to 1: In this case there
is always the opportunity of using \gem, provided the tensors can be generated
or stored in memory suitably. In this respect, the recipes also provide
guidelines on how tensors should be arranged for maximum performance.

If the user has no freedom in terms of generation and storage of the tensors,
\gem\ is still amenable, but at the cost of an overhead caused by copying and
transposing. Whenever possible, this option should be avoided.  Nonetheless,
in general \copgem\ still performs better than BLAS 2 kernels; moreover, the
transpositions can be optimized to reduce the number of memory accesses and
therefore the total overhead. 
In the next section, we present the performance
that one can expect for each of the classes.


\section{Numerical experiments}
\label{sec:sec5}

In this section, we present two numerical studies involving double
contractions.  The experiments mirror the two examples analyzed in
Sec.~\ref{sec:class3}; the aim is to support our deduction-based assumptions
on the use of BLAS 3, and give insights on which are the most
convenient alternatives when the direct use of \gem\ is ruled out. 

In the first study, we look at 3-dimensional
tensors with indices of equal size (square tensors), and measure the
performance as function of the size for four different alternatives---\gem,
\copgem, \gemv, \ger.

The second study is inspired by two common applications of tensorial
algebraic calculations: Coupled-Cluster method (CCD) in Quantum Chemistry, and
General Relativity on a lattice~\cite{TCE2, REL}.
In these applications, tensors are usually rectangular, with 
one of more order of magnitude of difference in the size of the indices.
Our objective is to verify that imposing the requirements of 
Sec.~\ref{sec:req} is still a valid option, even in extreme
cases in which the efficiency of BLAS 2 kernels is comparable to that of \gem{}

All the numerical tests were carried out on two different architectures: 
\tes, an 8-core machine based on the 2.00 GHz Intel Xeon E5405 processor
equipped with 16GB of memory (theoretical peak performance of 64 GFLOPS), and \pec, an 8-core machine based on the 
2.27 GHz Intel Xeon E5520 processor equipped with 24GB of memory (theoretical peak performance 72.6 GFLOPS).
In both platforms, we compiled with gcc (ver.~4.1.2) and used the 
OpenBLAS library\footnote{OpenBLAS is the direct successor of the GotoBLAS library~\cite{1377607}.}
(ver.~0.1) 
\cite{OBLAS1};
in all cases, double precision was used,
and performance is reported in terms of 
floating points operations/second.

\subsection{3-dimensional square tensors with $\Delta(t_i) =1$}
\begin{figure}[!tb] 
  \centering
  {\includegraphics[scale=0.17]{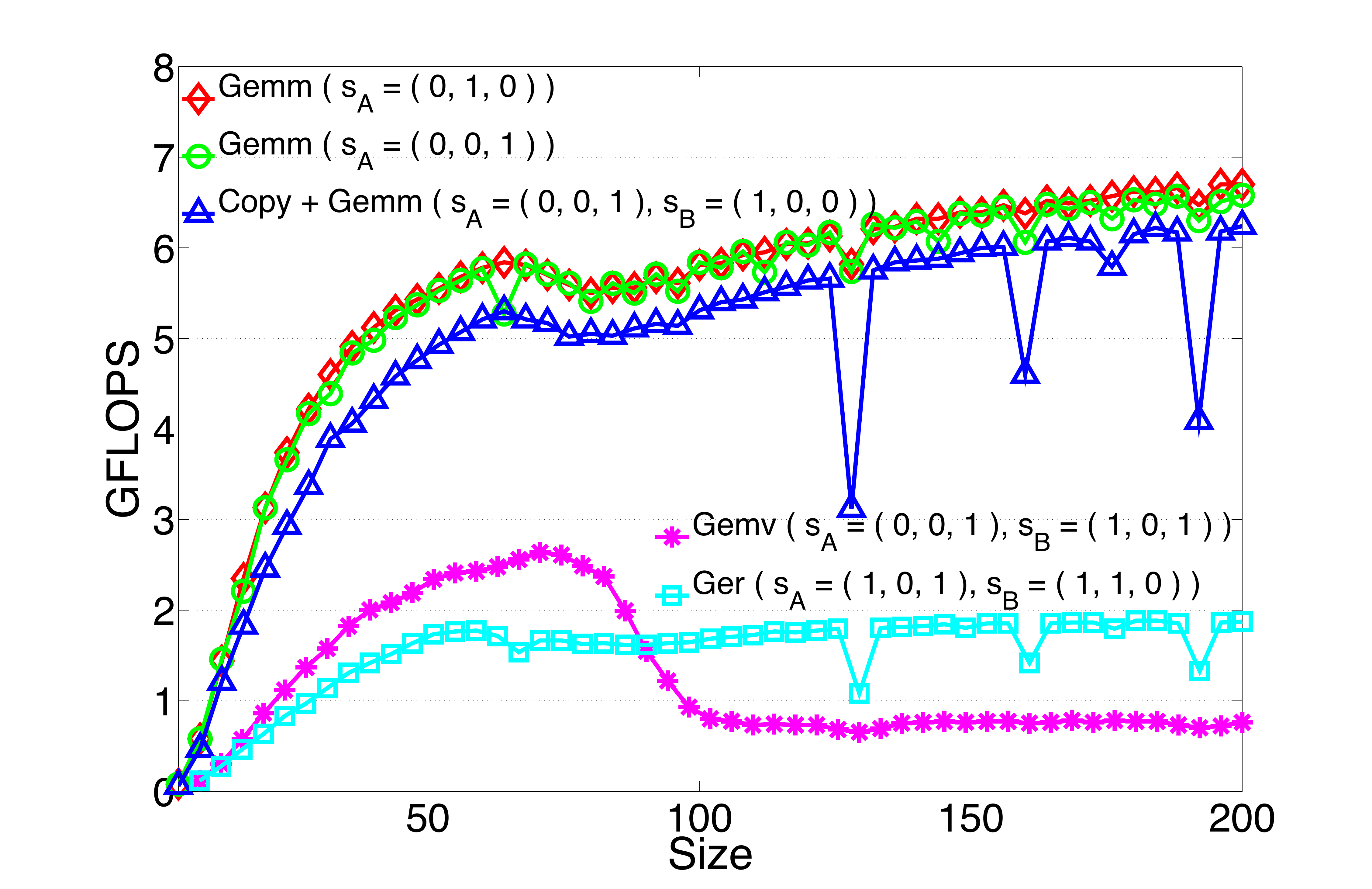}}
  \hspace{0.5cm}
  {\includegraphics[scale=0.17]{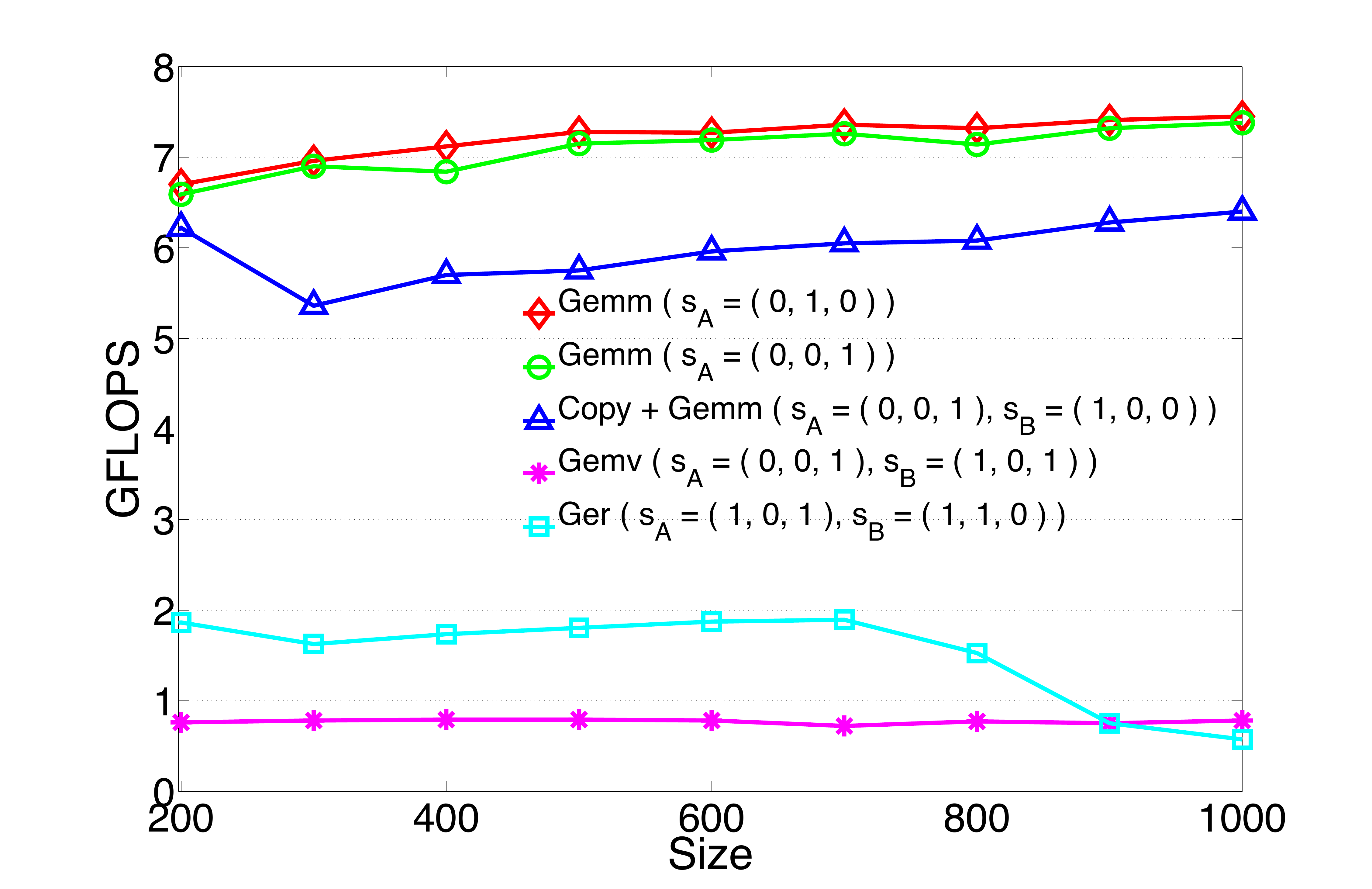}}
  \caption{Performance (number of floating point operations executed per second) 
    for the computation of 
    $R\indices{^\alpha_\eta} := A^{\alpha \beta \gamma} B_{\eta \beta \gamma}$ 
    and 
    $R\indices{^\beta_\eta} := A^{\alpha \beta \gamma} B_{\gamma \alpha \eta}$ 
    on 1 core of \tes.
  }
  \label{fig:3D-3}
\end{figure}

\begin{figure}[!tb] 
  \centering
  {\includegraphics[scale=0.18]{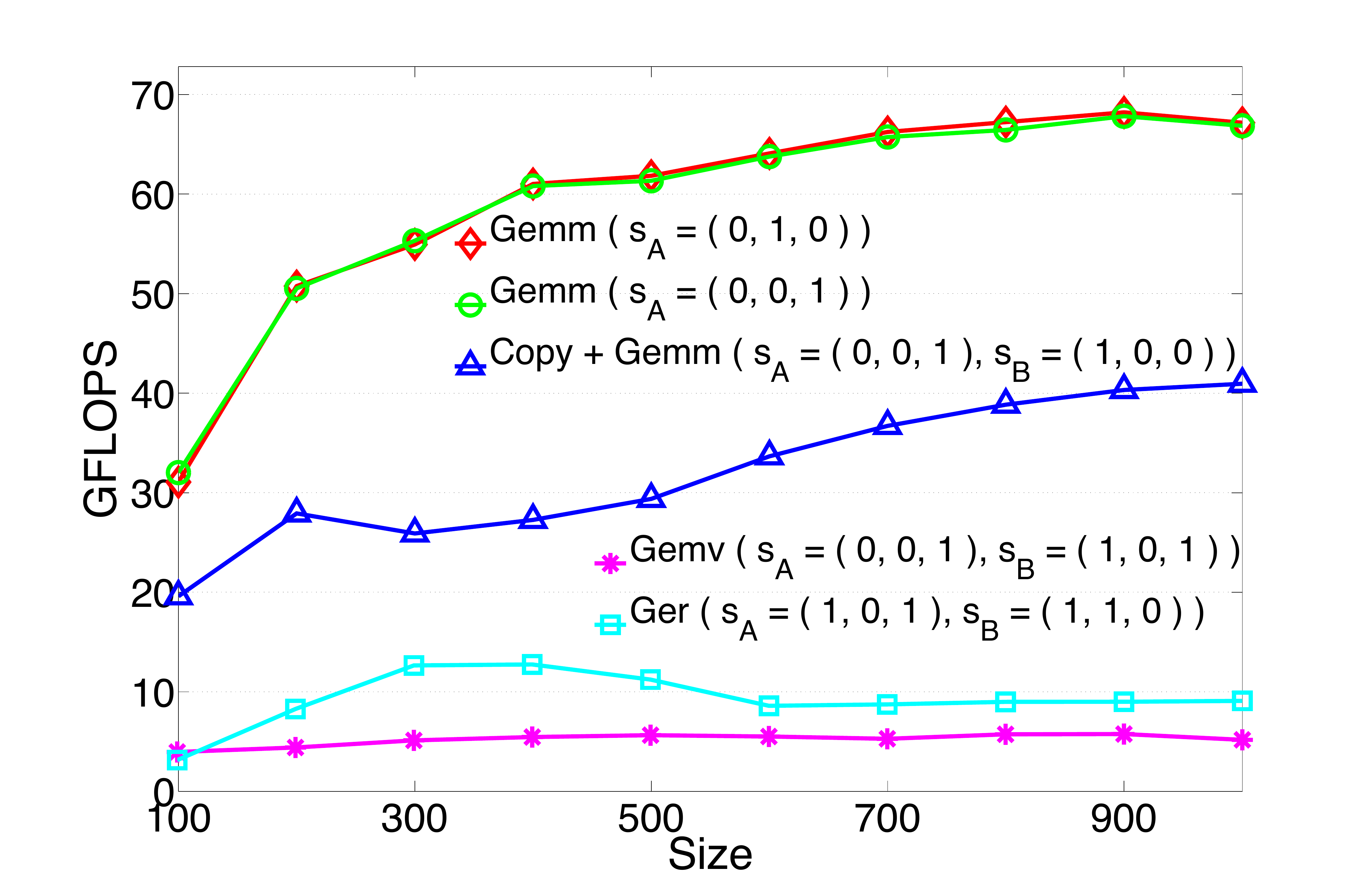}}
  {\includegraphics[scale=0.18]{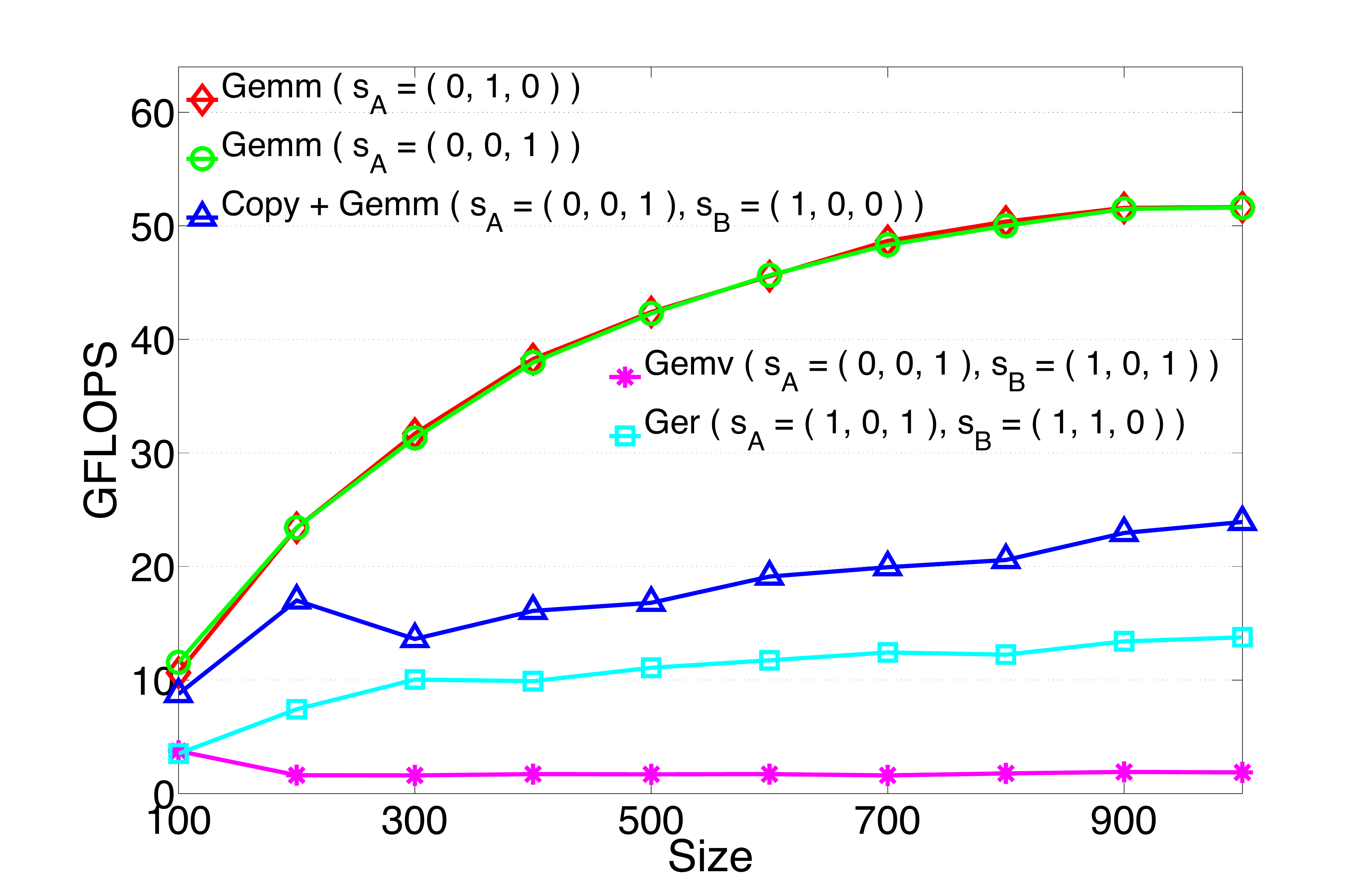}}
  \caption{
    Performance in GFLOPS for the computation of 
    $R\indices{^\alpha_\eta} := A^{\alpha \beta \gamma} B_{\eta \beta \gamma}$ 
    and 
    $R\indices{^\beta_\eta} := A^{\alpha \beta \gamma} B_{\gamma \alpha \eta}$ 
    on 8 cores of \pec (left) and \tes (right).
  }
  \label{fig:3D-4}
\end{figure}

We consider the contractions
$R\indices{^\alpha_\eta} := A^{\alpha \beta \gamma} B_{\eta \beta \gamma}$
and
$R\indices{^\beta_\eta} := A^{\alpha \beta \gamma} B_{\gamma \alpha \eta}$,
and the operations resulting from the slicings depicted in 
Fig.~\ref{fig:3D-1} and~\ref{fig:3D-2}.\footnote{
Since $A$ and $B$ are cubic tensors, 
these two contractions perform the same number of arithmetic operations. 
Moreover, if the tensors in the second contraction are stored in such a way that
a) $A$ is transposed so that the first and the second indices are exchanged,
and b) $B$ is transposed so that the first and the third indices are exchanged,
then they are also computing the same quantity, i.e.,
 $R\indices{^\alpha_\eta} = R\indices{^\beta_\eta}$.
}
Specifically, in Figg.~\ref{fig:3D-3} and~\ref{fig:3D-4}
we show the execution times for the computation of $R$
when using \gem\ (in two different ways), 
\copgem, \gemv, and \ger.

As expected, the two slicings leading to \gem\ outperform considerably those
that instead make use of BLAS 2 routines. 
In fact, for tensors of size 150
and larger, each routine attains the efficiency level mentioned in the
introduction: 90+\% of the theoretical peak for \gem{}, and between 10\% to
20\% of peak for BLAS 2 routines. 
This is an example of the clear benefits due to the use 
of BLAS-3 routines instead of those from BLAS 2 and 1. 
Additionally, the plots show that when \copgem{} is used,
for tensors of size 200 or larger
the overhead due to copying becomes apparent.\footnote{
  While it is possible to optimize the copying, amortizing the cost over
  multiple slices, here we present the timings for a simple slice-by-slice
  copy, as a typical user would implement.
} 
In short, the results are in agreement with
the discussion from Sec.~\ref{sec:stor}:
Whenever the contraction falls in the third class, 
the use of \gem\ is preferable. If the dimensions are small, there is little
difference between \copgem\ and \gem ; 
for mid and large sizes instead, 
it is always convenient to store the tensors so as to enable the direct use of \gem, 
thus avoiding any overhead (class 3.2).

Figure~\ref{fig:3D-4} illustrates the results for a parallel execution
on all the eight cores of \tes{}, using multi-threaded BLAS.
It is not surprising to observe the
increase of the gap between \gem{} and BLAS 2 based routines.
Moreover, comparing Fig.~\ref{fig:3D-3} and Fig.~\ref{fig:3D-4}, 
one can also appreciate the larger performance gap between \gem\ and
\copgem{}, confirming the fact that on multi-threaded architectures
it is even more crucial to carefully store the data to avoid unnecessary overheads.
In fact, as explained in Sec.~\ref{sec:stor}, the tensors with
$\Delta(t_i) \geq1$ 
should be stored in memory so that the dimension with stride 1
is not one of the contracted indices;
this choice prevents the use of \copgem\ and the 
consequent loss in performance.
Finally, we point out that even though 
the two \gem-based algorithms traverse the tensors in different 
ways, they attain the exact same efficiency. 

In conclusion, with this first example we confirmed
that when performing double contractions on square tensors,
any slicing that leads to \gem\  constitutes the most efficient choice,
The situation does not change as the order of the tensors and the number of
contracted indices increase. 

\subsection{4-dimensional rectangular tensors with $\Delta(t_i) =2$}
\label{sec:appl}
We shift the focus to real world scenarios in which the tensors are
rectangular and possibly highly skewed.  The objective is to test the
validity of our recipes in situations, inspired by scientific
applications, more general than contractions between square tensors.
For this purpose, we concentrate on double contractions between
4-dimensional tensors.  We present here two numerical examples: The
first arises in the Coupled-Cluster method, while the second is
motivated by covariant operations appearing in General Relativity. All
these tests refer to the tensors $X$ and $Y$ presented in
Sec.~\ref{sec:class3}, and the slicings described for
Eq.~(\ref{eq:xy}).  In particular, $R^{\alpha \beta \gamma \delta}$
was computed via four algorithms (resulting from four different
slicings), using \gem, \copgem, \dt, and \ger, respectively.%
\footnote{In this case the indices contracted remain the same,
  consequently $R^{\alpha \beta \gamma \delta} := X^{i \alpha j \beta}
  Y\indices{_i^\gamma_j^\delta}$ and $X^{i \alpha j \beta}
  Y\indices{_j^\gamma_i^\delta}$ involve exactly the same number of
  operations.  
}

\paragraph{Example 1: Coupled-Cluster method.}
For practical purposes, in Sec.~\ref{sec:class3} we introduced 
two different conventions for the tensor indices: 
We indicated the free and the contracted indices with letters from
the Greek and Latin alphabet, respectively;
when dealing with rectangular tensors, this notation is especially handy. 
In Quantum Chemistry, tensors indicate operators that allow the transition 
between occupied and virtual orbitals in a molecule. 
In the initial two examples, Greek indices correspond to occupied electronic 
orbitals, while Latin indices indicate unoccupied orbitals. 
Because of their physical meaning, 
the range of the indices $\alpha$, $\beta$, $\gamma$ and $\delta$ in 
Eq.~(\ref{eq:xy}) is quite different than that of $i$ and $j$.
Normally, the former is a group of ``short'' indices (size $\in [30, \dots,
100]$), while the latter are ``long'' indices 
(size $\in [1000, \dots, 3000]$)~\cite{TCE2}.

One of the consequences of these typical index ranges is that $X$ and
$Y$ can easily exceed the amount of main memory of standard multi-core
computers.  For example, if the indices $\alpha$, $\beta$, $\gamma$
and $\delta$ were all of size 80, and $i$ and $j$ were both of size
2000, the tensors $X$ and $Y$ would each require 204 GBs of RAM. In
practice contractions with such large tensors are executed in parallel
on clusters consisting of several computing nodes: $X$ and $Y$ are
then distributed among the memory of many multi-core nodes avoiding
altogether the issue of memory size.

\begin{figure}[!tb] 
  \centering
  \subfigure[Fixed $\alpha=\gamma=\delta=30$ and varying $i,j$]
  {\includegraphics[scale=0.19]{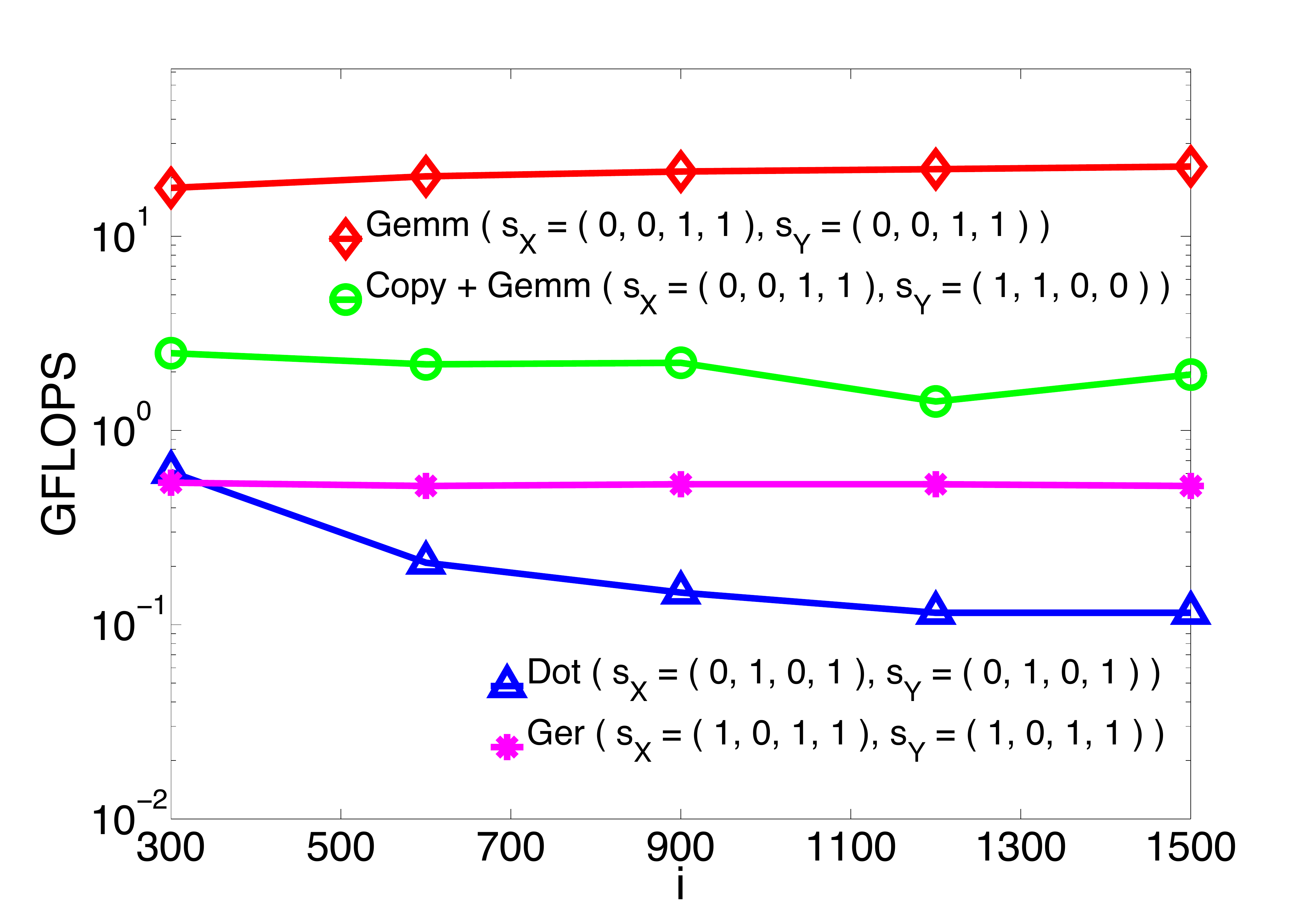}}
  \subfigure[Fixed $i=j=1,000$ and varying $\alpha ,\gamma ,\delta$]
  {\includegraphics[scale=0.19]{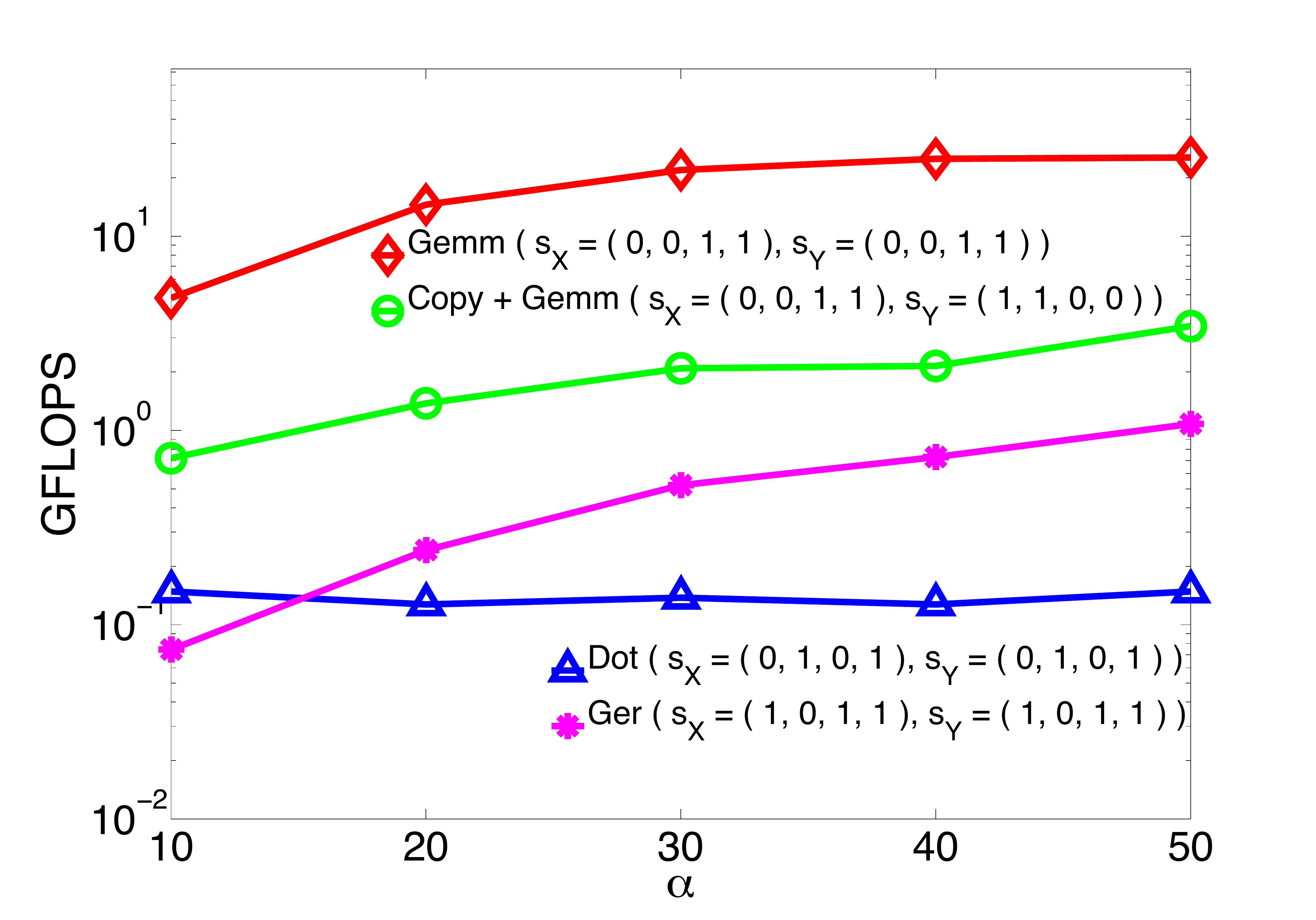}}
  \caption{
    Performance (in GFLOPS, on a logarithmic scale) for the computation of 
    $R^{\alpha \beta \gamma \delta} := X^{i \alpha j \beta} Y\indices{_i^\gamma_j^\delta}$
    on 8 cores of \pec. 
    On the left, the size of the indices $\alpha = \gamma = \delta$ is set to
    30, while that of $i = j$ varies. On the right, the size of $i = j$ is set
    to 1000, while that of $\alpha = \gamma = \delta$ varies. 
    In all cases, the free index $\beta$ is fixed.
  }
  \label{fig:4D-1}
\end{figure}

\begin{figure}[!tb] 
  \centering
  {\includegraphics[scale=0.19]{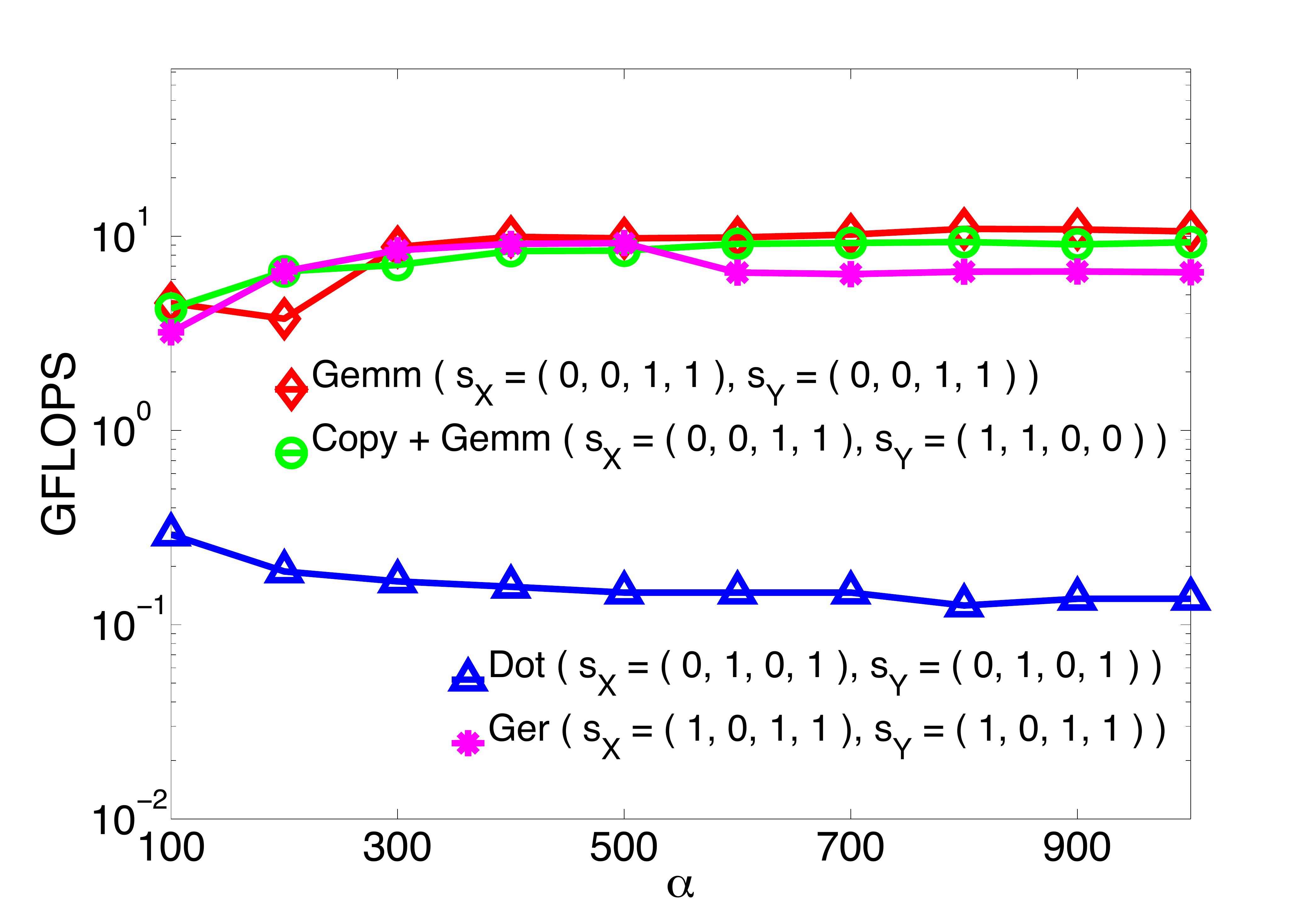}}
  {\includegraphics[scale=0.19]{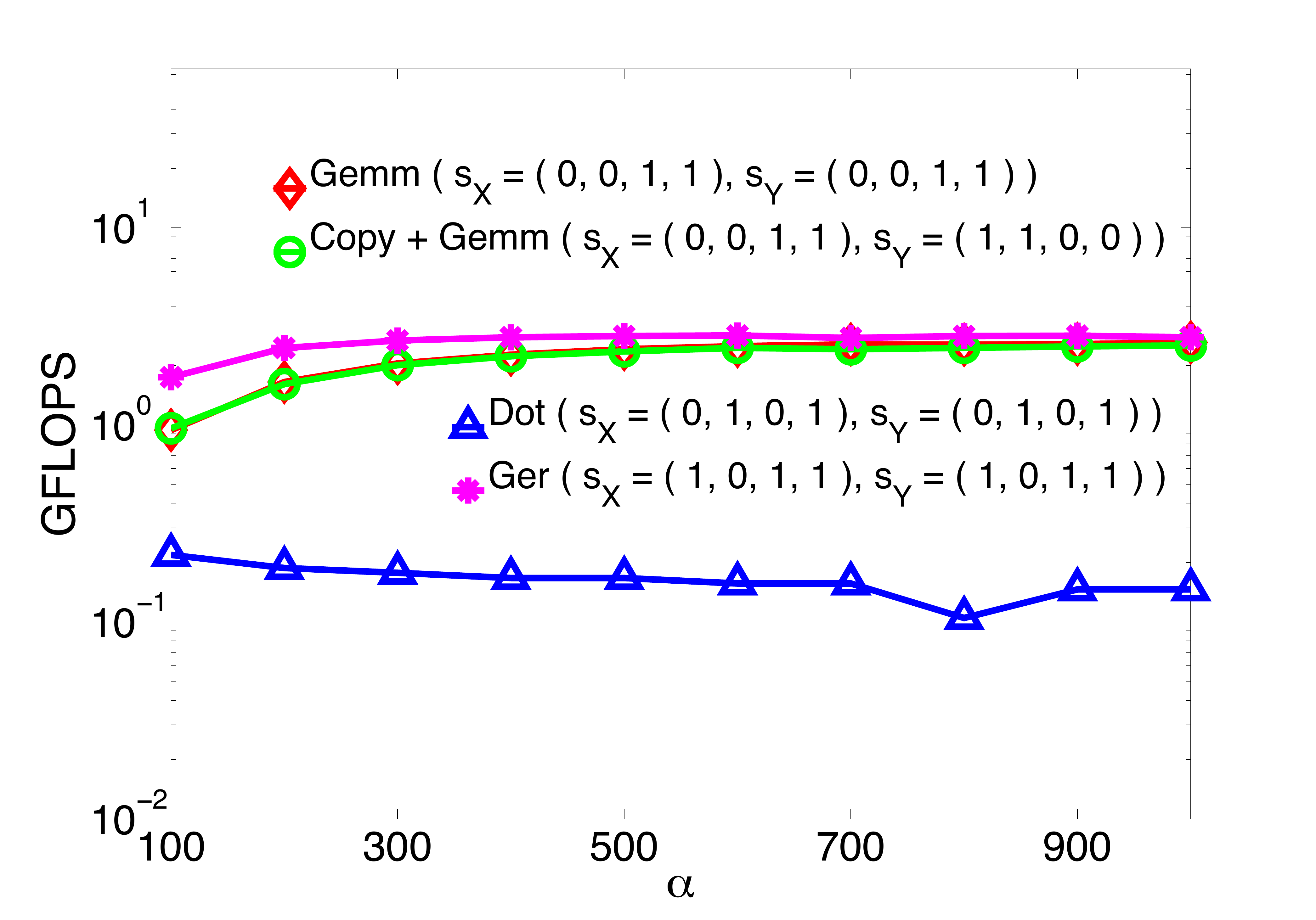}}
  \caption
  {
    Performance (in GFLOPS, on a logarithmic scale) for the computation of 
    $R^{\alpha \beta \gamma \delta} := X^{i \alpha j \beta} Y\indices{_i^\gamma_j^\delta}$
    on 8 cores of \pec (left) and \tes (right).
    The size of the indices $i$ and $j$ is set to 4, while 
    that of $\alpha = \gamma = \delta$ varies. 
    The free index $\beta$ is fixed.
  }
  \label{fig:4D-2}
\end{figure}

In order to run numerical tests on just one multi-core node, we
slightly reduced the size of the tensors, but kept the ratio between
the size of the long and short indices unchanged. Moreover due to the
heavy computational load, we fixed the free index $\beta$ to have
value equal to one (see Figg.~\ref{fig:4D-1} and
\ref{fig:4D-2}). Effectively, this is equivalent to computing one
single slice of $R^{\alpha \beta \gamma \delta}$ as opposed to all of
them; since $\beta$ is a free index, this simplification has no
influence over the performance of the algorithms.

The geometry of these tensors dictates that the 2-dimensional slices are not
square matrices anymore. In fact, the slicings that lead to \gem\ originate
long and skinny matrices with an aspect ratio of about 100:1; on the contrary,
those that lead to \ger\ originate many short vectors.  Figure~\ref{fig:4D-1}
(in logarithmic scale) suggests that even in these cases, the algorithms based
on \gem\ are preferable to those that instead make use of lower levels of
BLAS.  Indeed, in basically all tests, \gem\ is about one order of magnitude
faster than \copgem, and almost two orders faster than \ger\ and \dt.

\paragraph{Example 2: General relativity.}
In this last set of experiments, we consider the scenario in which the
contracted indices ($i, j$) are extremely short (size = 4), while the
free indices are long (size $\le$ 1,000).  This scenario takes
inspiration from a simplified version of 2-dimensional covariant
tensors discretized on a 4-dimensional pseudo-euclidean lattice
representing the space-time continuum.  In order to make calculations
and storage feasible, instead of working with 6-dimensional tensors,
we only operate on a subset of the indices, maintaining the number of
contractions unchanged.
Due to its highly non-square aspect ratio, this example represents a
rather extreme case that over-penalizes \gem\ with respect to the
other BLAS routines. In fact, while in the Coupled-Cluster example the
matrices were multiplied via \gem\ contracting their long side, here \gem\
contracts the skinny matrices along their short side. The same
scenario presents itself for the \dt\ case, where the contracted
slices are made of short vectors. On the contrary, the
slicings leading to \ger\ end up in outer-multiplications of long
vectors.

In Fig.~\ref{fig:4D-2} (in logarithmic scale), 
the performance of \gem\ drops substantially,
to the point that \gem\ becomes comparable to both \copgem\ and
\ger.  This sharp performance leveling is due to the extreme small
size of the contracted index of the input matrix slices: 
  While \ger{} attains its peak performance due to the large size
  of the involved vectors, \gem{} is rarely optimized for this specific
  shape (short contracted index).
In the case of \tes, the \ger-based contractions are, even
if slightly, more efficient than \gem.  Although this result is
somewhat expected, it shows that even in this example \gem-based
contractions are at least at the same level with the other BLAS
sub-routines. \dt\ performance is once again lower than the other
routines confirming that even in this extreme case it is always best
to use higher levels of BLAS.

In conclusion, all the tests confirm the
prominent role of \gem\ for algebraic tensor operations and establish
the necessity of adopting tensor storing practices that favor 
its use.

\section{Conclusions} \label{sec:conclusions} 

In this paper we look at multi-contractions between two tensors
through the lens of the highly optimized BLAS libraries. Since the BLAS routines
operate only with vectors and matrices, we
illustrated a slicing
technique that allows the structured reduction of each tensor
into many lower dimensional objects. 
Combined with constraints of the BLAS interface, 
this technique naturally leads to the formulation of a
set of requirements for the use of \gem, the best performing 
BLAS routine. By using such requirements, we carried out a systematic
analysis which resulted in a classification of tensor contractions in
three classes. For each of these classes we provided practical guidelines
for the best use of the available BLAS routines.

In the last section numerical experiments were performed with the aim
of testing the validity of our guidelines with particular attention to
real world scenarios. All examples confirm that for the optimal use
of the BLAS library, particular care has to be placed in generating and storing
the tensor operands. Specifically for contractions leaving at least one free index
per tensor, the operands should be directly initialized and allocated
in memory so as to avoid the index with stride 1 be the
contracted one, and in doing so favor the direct use of \gem.

\newpage
\begin{flushleft}
      {\bf \Large Appendix}
    \end{flushleft}
\begin{appendices}
\numberwithin{equation}{section}
\section{Tensors on a manifold}
\label{Appendix}
The scientific community generically refers to multi-dimensional arrays as
tensors, and uses this term in its most broadly defined
sense. Sometimes such a fuzzy use of the term ``tensor'' may create
confusion in the reader that is novel to this field. To avoid any
confusion we will refer, in the following, to the mathematically
rigorous definition of tensor used in differential geometry (sometimes
also called tensor field). In this context a tensor is a mathematical
object whose properties depend on the point of the geometrical
manifold where it is defined. If such a manifold is a simple euclidean space
provided with cartesian coordinates, a tensor is indistinguishable
from a multi-dimensional array. On the contrary, adopting a more
stringent tensor definition, allows to extend the results of this
paper to the most general case of tensor fields.

The geometrical properties of a manifold $X$ are, in general, encoded
in a symmetric and non-singular (admitting an inverse) two-index
object $g_{\mu \nu}$ called ``metric''. In a manifold there are two intrisically
different objects that have the characteristics of a traditional
vector: the vector proper $V^\mu$ and the dual vector $\omega_\mu$. A vector is changed into a dual vector by the metric
\[
V_{\mu'} = \sum_{\mu =1}^N V^\mu g_{\mu \mu'} \equiv V^\mu g_{\mu \mu'},
\]
where the {\sl Einstein} convention has been used in the last
equality. Due to the uniqueness of the metric in defining the geometry
of $X$, scalar products can only be computed among vectors once one of
them is transformed into a dual vector by the metric
\[
V^\mu \cdot W^\nu = V^\mu g_{\mu \nu} W^\nu \equiv V^\mu W_\mu \equiv V_\nu W^\nu .
\]
This is what we refer to as ``index contraction''.

The above expressions can be interpreted in the language of linear
algebra if we think of a vector $V^{\mu}$ as a column n-tuple and a
dual vector as a row n-tuple
\[
	V^{\mu} = \left(
				\begin{array}{l}
					V^1 \\
					\vdots \\
					V^n \nn
				\end{array}
	\right) \qquad ; \qquad \omega_{\mu} = (\omega_1 \dots \omega_n)
\]
Based on this analogy we can establish a notational language by
imposing that
\begin{itemize} \centering
	\item[] any LOWER (covariant) index $\longrightarrow$  is a COLUMN index
	\item[] any UPPER (contravariant) index $\longrightarrow$ is a ROW index
\end{itemize}
So a general multi-indexed tensor with $M$ lower and $N$ upper indices
$T\indices{^{\mu_1' \dots \mu_N'}_{\nu_1' \dots \nu_M'}}$ can be
thought as a multi-array object having
\begin{itemize} \centering
	\item[] M covariant indices corresponding to M \textbf{column-modes}
	\item[] N contravariant indices corresponding to N \textbf{row-modes}.
\end{itemize}

If with $g_{\mu \nu}$ one can lower the indices of a generic
tensor, the inverse of the metric $g^{\mu \nu}$ is used to raise tensor
indices. In other words the metric transforms a row index in a column
index while the inverse of the metric accomplishes the opposite. For
example
	\begin{eqnarray*}
		T\indices{^{\alpha \beta \gamma}_{\rho \sigma \tau}} g_{\beta \nu} = & T\indices{^{\alpha}_{\nu}^{\gamma}_{\rho \sigma \tau}}\\
		T\indices{^{\alpha \beta \gamma}_{\rho \sigma \tau}} g^{\sigma \mu} = & T\indices{^{\alpha \beta \gamma}_{\rho}^{\mu}_{\tau}}\\
	\end{eqnarray*} 
From these examples we can infer the following:
\bi
      \item[1.] lowering is equivalent to row-mode multiplication with
        any (since $g$ is symmetric) column mode of the metric;
      \item[2.] raising is equivalent to column-mode multiplication
        with any row-mode of $g$. 
\ei

Having defined the basic rules and language, we can now define a
generic multiple contraction between two tensors 
\[
T\indices{^{\alpha \beta \gamma}_{\rho \sigma \tau}} g_{\beta \eta}
g^{\sigma \xi} S\indices{^{\mu \eta}_{\xi \nu}} 
= T\indices{^{\alpha \beta \gamma}_{\rho \sigma \tau}}
S\indices{^{\mu}_{\beta}^\sigma_\nu}
= R\indices{^{\alpha \mu \gamma}_{\rho \beta \tau \nu}}.
\]
In other words a contraction between two tensors $T$ and $S$ requires:
\bi
\item[i.] two indices carrying the same symbol (repeated indices), one
  on $T$ and one on $S$;
\item[ii.] one of the two repeated indices has to be a row mode while the other
  has to necessarily be a column mode;
\item[iii.] each lowering and rising operation is performed by a metric tensor.
\ei

As for matrices, a tensor can be characterized by explicitly
indicating its indices as well as their respective order. With these
convention, a common matrix is a 2-tensor in a cartesian euclidean
space. A general 2-tensor is not a matrix though, and matrix-matrix
multiplications depend strictly on the geometry of the manifold where
such 2-tensor live.  In euclidean space with a cartesian coordinate
system, the metric is just the identity but already if the space is
equipped with spherical coordinates this is not true anymore

For instance, if we want to describe $\mathbb{R}^3 - \{0\}$ in spherical coordinates it is necessary to make the following change of coordinates
\[
\left(x,y,z\right) \overset{f}{\longrightarrow} \left(r,\theta,\phi
\right) : \qquad f^{-1} = \left\{
	\begin{array}{lcl}
	x=r \sin(\theta)\cos(\phi) & {\rm for} & r \geq 0\\
	y=r \sin(\theta)\sin(\phi) & {\rm for} & 0\leq \theta < \pi\\
	z=r \cos(\theta) & {\rm for} & 0\leq \phi < 2\pi .
	\end{array}
	\right.
\]
Such coordinate transformation induces a metric that is not anymore
proportional to the identity even if it retains its diagonal form
\[
	\left(
	\begin{array}{ccc}
	1 & 0 & 0 \\
	0 & r^2 & 0 \\
	0 & 0 & r^2\sin^2(\theta)
	\end{array}
	\right).
\]
In this simple euclidean manifold lowering the indices of a
(2,0)-tensor defined at a certain point of the manifold produces the
following result
\[
	S^{\mu\nu} = \left(
	\begin{array}{ccc}
	S_{11} & S_{12} & S_{13}\\
	S_{21} & S_{22} & S_{23}\\
	S_{31} & S_{32} & S_{33}
	\end{array}
	\right)
	\quad \longrightarrow \quad
	S_{\mu\nu} = \left(
	\begin{array}{ccc}
	S_{11} & r^2S_{12} & r^2\sin^2(\theta)S_{13}\\
	r^2S_{21} & r^4S_{22} & r^4\sin^2(\theta)S_{23}\\
	r^2\sin^2(\theta)S_{31} & r^4\sin^2(\theta)S_{32} & r^4\sin^4(\theta)S_{33}
	\end{array}
	\right)
\]
This simple example shows the importance of keeping contravariant and
covariant indices quite distinguished.

\end{appendices}
